\documentclass[acmsmall]{acmart}

\usepackage{amsmath}
\usepackage{amsfonts}
\usepackage{mathtools}
\usepackage{multirow}
\usepackage{subfigure}
\usepackage{adjustbox}
\usepackage{wrapfig}
\usepackage{xspace}
\usepackage{algorithm}
\usepackage{algpseudocode}
\usepackage{csquotes}
\usepackage{booktabs}
\usepackage{graphicx}
\usepackage{epstopdf}
\usepackage{bm}
\usepackage{hyperref}

\newcommand{\eg}{\emph{e.g.},\xspace}
\newcommand{\tb}[1]{\textbf{#1}}

\usepackage{graphicx}
\usepackage{array}

\setcopyright{acmcopyright}
\copyrightyear{2023}
\acmYear{2023}
\acmDOI{XXXXXX}

\begin{document}

\title{A Comprehensive Survey on Multimodal Recommender Systems: Taxonomy, Evaluation, and Future Directions}

\author{Hongyu Zhou}
\email{hongyu.zhou@ntu.edu.sg}
\affiliation{
  \institution{Nanyang Technological University}
  \country{Singapore}
}

\author{Xin Zhou}
\email{xin.zhou@ntu.edu.sg}
\affiliation{%
  \institution{Nanyang Technological University}
  \country{Singapore}}

\author{Zhiwei Zeng}
\email{zhiwei.zeng@ntu.edu.sg}
\affiliation{%
  \institution{Nanyang Technological University}
  \country{Singapore}}

\author{Lingzi Zhang}
\email{lingzi001@e.ntu.edu.sg}
\affiliation{%
  \institution{Nanyang Technological University}
  \country{Singapore}}

\author{Zhiqi Shen}
\email{ZQShen@ntu.edu.sg}
\affiliation{%
  \institution{Nanyang Technological University}
  \country{Singapore}}

\begin{abstract}
Recommendation systems have become popular and effective tools to help users discover their interesting items by modeling the user preference and item property based on implicit interactions (\eg purchasing and clicking). Humans perceive the world by processing the modality signals (\eg audio, text and image), which inspired researchers to build a recommender system that can understand and interpret data from different modalities. Those models could capture the hidden relations between different modalities and possibly recover the complementary information which can not be captured by a uni-modal approach and implicit interactions. The goal of this survey is to provide a comprehensive review of the recent research efforts on the multimodal recommendation. Specifically, it shows a clear pipeline with commonly used techniques in each step and classifies the models by the methods used. Additionally, a code framework has been designed that helps researchers new in this area to understand the principles and techniques, and easily runs the SOTA models. Our framework is located at: \url{https://github.com/enoche/MMRec}.

\end{abstract}



\maketitle

\section{Introduction}

With the development of society, a huge number of items and information have been displayed in the network, it would be difficult for users to decide which is useful and pick out the preferred one. The information overload issue has been solved when the recommendation system appears\cite{lu2012recommender}. It predicts the rating or user preference over all items and recommends the most probable and pertinent items that the user might choose according to the historical interaction data and public item information(like clicks and sales). The algorithms of recommendation could be classified into collaborative filtering (CF), content-based filtering and hybrid recommendation system~\cite{shah2016survey}. CF is based on analyzing and gathering the user's historical behaviors data, which includes the historical interactions (\eg clicks, look-through, purchases) and user preference (\eg ratings). The content-based filtering suggests products based on the user profile and item profile of the user. The item is described with the keywords and the user's profile will express the types of items the user likes. The main idea of this method is the user will probably choose similar items that they like before. Hybrid approach combines different techniques of collaborative filtering and content-based filtering to get a better result.

Those traditional recommendation methods~\cite{he2020lightgcn,zhou2021selfcf,zhang2022diffusion,zhou2022layer} have shortages as we introduce that they require a large number of interactions between users and items to make a more accurate recommendation. The users and items with few or even no interactions will influence the accuracy of recommendations. 
In order to alleviate the data sparsity problem~\cite{zhou2016evaluating,zhou2022bribery} and cold start issue, multimodal information has been introduced into the recommendation system. The multimodal recommendation leverages the auxiliary multimodal information which could supplement the historical user-item interactions to improve the recommendation performance. The multimodal model is able to represent and discover hidden relations between different modalities and possibly recover the complementary information which can not be captured by a uni-modal approach and implicit interactions. Such skills are also necessary for natural language processing~\cite{sulubacak2019multimodal} to achieve human-level comprehension in a variety of AI jobs. The multimedia data shows not only the relationship between users and items but also reflect the preference of user in different modalities. For the same modality, the data might reflect the similarity and the semantic meaning of different items. In order to incorporate multimodal information into the recommendation system, the current method is extracting the features from different modalities and then using the modality fusion result as the side information or the item representation. VBPR~\cite{he2016vbpr} is the first model that considers introducing visual features into the recommendation system by concatenating visual embeddings with id embeddings as the item representation. \cite{wei2019mmgcn,wang2021dualgnn,wei2020graph} utilize the GCN-based methods to produce the representations of each modality and then fuse them together as the final representations. Except for fusing the modality representations, the knowledge graph based modality side information is also introduced into the multimodal recommendation~\cite{sun2020multi}. Different techniques have been introduced into the multimodal recommendation that tries to find better representations for users and items to achieve more accurate recommendation results.

Nowadays, online sharing platforms are commonly used, such as fashion, news, short video and music recommendation platforms. Those popular online platforms contain a large number of multimodal information that drives the user's choice. As people commonly use those platforms with huge multimodal information contained inside, the multimodal recommendation is necessarily applied to learn more accurate user preferences.
Different from the traditional recommendation, those applications have utilized the item multimodal content information like the video frames, audio track and item descriptions. MMGCN~\cite{wei2019mmgcn}, MGAT~\cite{tao2020mgat}, DualGNN~\cite{wang2021dualgnn} and SLMRec~\cite{tao2022self} are the micro video recommendation models that utilize description, captions, audio, and frames inside the video to model the multimodal user preference on the micro-video. Fashion recommendation faces difficulties to build an efficient recommender cause of the complex features involved and the subjectivity. \cite{kang2017visually,chen2019personalized} aims to utilize multimodal information to overcome the issue. The news recommendation previously focuses on the title and text uni-modal information but ignores the visual information like images. \cite{wu2021mm,xun2021we} try to use multi-modality to enhance the recommendation performance. 

The multimodal recommendation system is popularly used because of the advantages and necessities we discussed above. And there is no survey that discusses the techniques used for multimodal recommendation. In order to provide a guideline for researchers, we describe the overview of the multimodal recommendation system and classify the reviewed papers according to the methods used. The goal of this survey is to introduce how the multimodal recommendation system works by showing a clear pipeline of the process with techniques used in each step. We aim to guide new researchers to understand the principles and techniques commonly used in multimodal recommendation systems. Additionally, we provide a developed code framework that contains the implemented SOTA models with pre-extracted multimodal features, which makes it efficient for researchers to run the SOTA models or develop their own model based on the framework.

\subsection{Strategy for Literature search}
Since our survey focuses on reviewing the multimodal recommendation, we retrieved most of the top conferences related to recommendation systems such as WWW, SIGIR, KDD, ICLR, AAAI, WSDM, and RecSys, also the top journals such as TKDD, TMM, and so on. At the same time, we also utilize Google Scholar to search for related papers. We use the keywords multimodal recommendation, multimodal recommendation, multimedia recommendation, content+RS, context+RS, etc, to search for the related papers. After getting the list of papers, we also look through the related works of those papers to make sure that we covered a majority of papers about multimodal recommendations.

\subsection{Differences compared with other surveys}
There are surveys that have looked into the recommendation area because of the importance and popularity of the recommendation system ~\cite{le2022survey,wang2021graph,shi2014collaborative,rafsanjani2013recommendation}. To the best of our knowledge, those surveys do not consider the multi-modality information and there is only one survey~\cite{deldjoo2020recommender} that discusses the multimedia recommendation and mentions the multi-modality information. Although discussion of multi-modality always involves both modality and media, these two terms are not synonymous but with overlapping~\cite{lauer2009contending}. The media is a physical channel that uses technology to store, move, and transmit information. It is the tool and material resource to produce and disseminate information. The modality is a human sensory channel that serves as the user input modality when interacting with a system. It can be considered a way of representing information. For one media, there might be several modalities contained inside. For example, books are considered to be media but the different languages of text are considered as modalities.  Additionally, this survey only aims to bridge the recommendation system with multimedia and just classify the papers according to the media type and application. It considers which application those modalities could be used for (like music recommendation, movie recommendation, etc.) but ignores the hidden algorithm and methods that are used to utilize the multi-modality information in the recommendation system, and also ignores the fusion methods of different modalities.

Unlike the existing surveys, our survey is organized according to the different methods usually used for multimodal recommendation systems. The mentioned multimedia survey just wants to divide models according to the media type. What we want to achieve is that the new beginners could learn what are the main steps of the multimodal recommendation system, what models could be used for their purpose and what are the mainly used techniques in each step like extracting features from different modalities and fusing the different modalities' representations. Another difference in our survey is that we not only provide the summaries of those multimodal recommendation models but develop a framework that contains the SOTA models utilizing different methods with several commonly used datasets. The beginners could easily run those baseline models after reading this survey, and they could conveniently develop their model with add new datasets according to the format.

\subsection{Organization and Contribution of Survey}
The main content of this survey is the overview of research on multimodal recommendation. The purpose of our work is to summarize the main methods that are used in this area and highlight the importance of multimodal content. Also, we will provide a common framework that contains the code implementation of the commonly used models which could help the beginner quickly run their data or implement the models in this framework. We will organize the survey as follows:
\begin{itemize}
    \item Introduction 
    \item Pipeline of multimodal recommendation
    \item Feature Extraction
    \item Model Classification
    \item Modality Fusion
    \item Measurement and Optimization
    \item Dataset and experiment results
    \item Challenges and Future research directions
    \item Conclusion
\end{itemize}

Having presented the organization of this survey, we summarize our main contributions as follows.

\begin{itemize}
\item We present a theoretical review of the multimodal recommendation system (MMRec), outline the various strategies for learning and leveraging modality information, and provide a comprehensive explanation of the techniques the model utilizes and how each MMRec model works.
\item We present the main steps of how the recommendation system leverages multimodal information and shows the pipeline of MMRec. Discuss how to choose the suitable dataset, feature extraction and optimization methods, which provide a clear guide to readers on developing the MMRec system.
\item We validate the effectiveness and efficiency of the SOTA multimodal recommendation models by performing extensive experiments on four public datasets. Also, discuss the key factor that influences models' performance (\eg different modality information, different data split methods).
\item Additionally, we create an open-source framework, MMRec~\cite{zhou2023mmrecsm}\footnote{https://github.com/enoche/MMRec}, that could be used to implement the baselines of MMRec models mentioned in this survey. Readers could also use this code framework to develop their own models. We hope this framework helps beginners easier understand how the multimodal recommendation model works and conveniently develop their own model considering only how to design the model with other steps done by the framework. In addition, we created an open-source repository that includes all those reviewed papers and the unimodal models\footnote{https://github.com/hongyurain/Recommendation-with-modality-information}.
\end{itemize}

\section{Pipeline of Multimodal Recommendation}
\begin{figure}[H]
\centering
\includegraphics[scale=0.4]{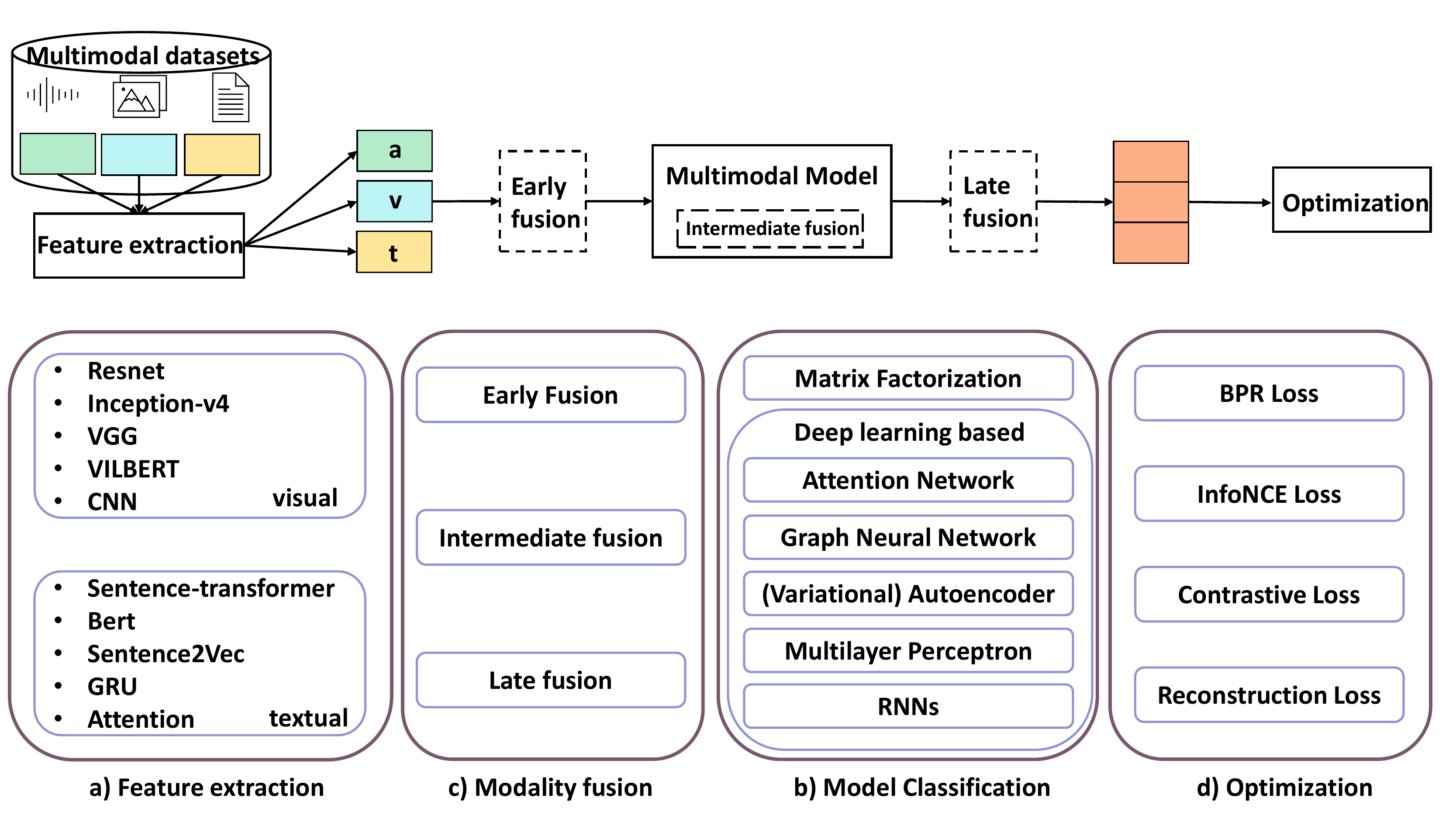}
\caption{Pipeline of multimodal recommendation}
\label{Pipeline}
\end{figure}

The multimodal recommendation system generally follows the steps in Fig.~\ref{Pipeline} and the modality fusion methods will be in different steps according to the settings which have been squared by dotted lines in the graph. The details of each step have been described in the following.

The multimodal recommendation system aims to learn informative representations of users and items by leveraging multimodal features. The first step is to extract modality features from the raw data. After getting the modality feature, you can choose to fuse them before feeding them into the multimodal models, fuse them in the intermediate layer of the model, or fuse the output of multimodal models. There will be different techniques applied to learn the user-item relations and leverage the multimodal information, which improves the recommendation accuracy with matched objective functions.

\section{Feature Extraction}

The commonly used datasets may not contain all the modality features so it is also important to select suitable datasets according to the modality included. Table~\ref{tab: common dataset} shows the most commonly used datasets are the Amazon dataset, Movielens, Tiktok and Kwai. The Amazon dataset always contains image and text information, and there are many categories under Amazon so the selection of the dataset will also depend on the category. Baby, Sports, Beauty and Arts are commonly used for multimodal recommendation~\cite{zhou2022bootstrap,zhang_mining_2021,zhou2023enhancing}. And for some special domains like fashion recommendation, it always utilizes the Men and Women categories under Amazon. Food recommendations use the Yelp dataset which contains images and user reviews. News recommendations use the MIND dataset. The Tiktok, Kwai and Movielens datasets are used for the short video recommendation which contains textual, visual and audio information~\cite{wei2019mmgcn,wang2021dualgnn}, but Tiktok and Kwai are not publicly available. If you need special modality information or there are no datasets for your research domain, you can also crawl the data from the website and existing datasets to establish a new dataset.

\begin{table}[h]
\centering
\renewcommand\arraystretch{1.5}{
\begin{tabular}{m{9em}m{28em}}
\toprule
Dataset              & Models                \\
\hline
Amazon    & VBPR\cite{he2016vbpr}, VMCF\cite{park2017also}, AMR\cite{tang2019adversarial}, VECF\cite{chen2019personalized}, PAMD\cite{han2022modality}, PMGT\cite{liu2021pre}, LATTICE\cite{zhang_mining_2021}, JRL\cite{zhang2017joint}, BM3\cite{zhou2022bootstrap}, GraphCAR\cite{xu2018graphcar}, DVBPR\cite{kang2017visually}, MAML\cite{liu2019userdiverse}, DMRL\cite{liu2022disentangled}, MVGAE\cite{yi2021multi}, MMCPR\cite{liu2022multi}, HCGCN\cite{mu2022learning}, FREEDOM~\cite{zhou2022tale}, DRAGON~\cite{zhou2023enhancing} \\
\hline
Kwai& MMGCN\cite{wei2019mmgcn}, SLMRec\cite{tao2022self}, GRCN\cite{wei2020graph}, EliMRec\cite{liu2022elimrec}, InvRL\cite{du2022invariant}, A2BM2GL\cite{cai2022adaptive} \\
\hline
Tiktok& MMGCN\cite{wei2019mmgcn}, DualGNN\cite{wang2021dualgnn}, MGAT\cite{tao2020mgat}, SLMRec\cite{tao2022self}, GRCN\cite{wei2020graph}, MMGCL\cite{yi2022multi}, EgoGCN\cite{chen2022breaking}, EliMRec\cite{liu2022elimrec}, InvRL\cite{du2022invariant}, A2BM2GL\cite{cai2022adaptive}, LUDP\cite{lei2022learning}\\   
\hline
MovieLens& MMGCN\cite{wei2019mmgcn}, DualGNN\cite{wang2021dualgnn}, MGAT\cite{tao2020mgat}, SLMRec\cite{tao2022self}, GRCN\cite{wei2020graph}, MMGCL\cite{yi2022multi}, MKGAT\cite{sun2020multi}, EgoGCN\cite{chen2022breaking}, EliMRec\cite{liu2022elimrec}, InvRL\cite{du2022invariant}, A2BM2GL\cite{cai2022adaptive}, LUDP\cite{lei2022learning}\\
\hline
Yelp& PAMD\cite{han2022modality}\\
\hline
MIND& IMRec\cite{xun2021we} \\
\hline
Dianping& MKGAT\cite{sun2020multi} \\
\hline
Alishop& MVGAE\cite{yi2021multi} \\
\hline
Pinterest& ACF\cite{chen2017attentive}, AMR\cite{tang2019adversarial} \\
\hline
Crawl dataset& PAMD\cite{han2022modality}, MM-Rec\cite{wu2021mm}, UVCAN\cite{liu2019user}, DVBPR\cite{kang2017visually}, ACF\cite{chen2017attentive} \\
\bottomrule
\end{tabular}}
\caption{Commonly used Dataset}
\label{tab: common dataset}
\end{table}

The purpose of feature extraction is to describe the modality features in a low-dimensional and explainable way with embeddings. The modality data can be utilized in two ways. The first is passing the pre-extracted modality data into the model, and the second is passing the raw data into the model by applying end-to-end learning. According to most of the MMRec model settings, our framework are following the first method to use pre-extracted features. Different modality data have modality-specific methods to extract. Table~\ref{feature extraction} summarize the multimodal feature extraction methods. Depending on the different modalities, the data resources and methods are different. We simply divided the task according to the data resources and list out the mainly used extraction methods with the example models. 

We can figure out that for visual features, the commonly used techniques are the CNN and variants of CNN. For the models~\cite{wei2019mmgcn,wang2021dualgnn} utilize Kwai and Tiktok, Those two datasets are not publicly available, even the researchers that own the datasets utilize the visual and textual features which are pre-extracted. For the textual feature, different pre-trained language models are used according to the purpose of the model.

\begin{table}
\centering
\begin{tabular}{m{6em} m{6em} m{7em} m{18em}}
\toprule
Modality                 & Content                  & Methods      & Models \\
\hline
Visual                  &Micro Video         & Published \qquad feature   &MMGCN\cite{wei2019mmgcn}, DualGNN\cite{wang2021dualgnn}, MGAT\cite{tao2020mgat}, SLMRec\cite{tao2022self}, GRCN\cite{wei2020graph}, HHFAN\cite{cai2021heterogeneous}, MMGCL\cite{yi2022multi}, EgoGCN\cite{chen2022breaking}, InvRL\cite{du2022invariant}, A2BM2GL\cite{cai2022adaptive}, LUDP\cite{lei2022learning}        \\
\hline
\multirow{7}{*}{Visual}  & \multirow{7}{*}{Images}  & VGG-19       & VECF\cite{chen2019personalized}, PAMD\cite{han2022modality}       \\
                         &                          & VILBERT      & MM-Rec\cite{wu2021mm}       \\
                         &                          & CNN+MLPs     & IMRec\cite{xun2021we}       \\
                         &                          & Inception-v4 & PMGT\cite{liu2021pre}       \\
                         &                          & Inception-v3 & UVCAN\cite{liu2019user}       \\
                         &                          & Deep CNN     & VBPR\cite{he2016vbpr}, VMCF\cite{park2017also}, LATTICE\cite{zhang_mining_2021}, FREEDOM~\cite{zhou2022tale}, HCGCN\cite{mu2022learning}, JRL\cite{zhang2017joint}, BM3\cite{zhou2022bootstrap}, GraphCAR\cite{xu2018graphcar}, DVBPR\cite{kang2017visually}, DRAGON~\cite{zhou2023enhancing}       \\
                         &                          & Caffe        & MAML\cite{liu2019userdiverse},DMRL\cite{liu2022disentangled}       \\
                         &                          & ResNet     & MKGAT\cite{sun2020multi}, MVGAE\cite{yi2021multi}, GraphCAR\cite{xu2018graphcar}, ACF\cite{chen2017attentive}, AMR\cite{tang2019adversarial}, EgoGCN\cite{chen2022breaking}, EliMRec\cite{liu2022elimrec}, InvRL\cite{du2022invariant}     \\
\hline
\hline
Textual                  & Video Titles             & Published \qquad feature   & MMGCN\cite{wei2019mmgcn}, DualGNN\cite{wang2021dualgnn}, MGAT\cite{tao2020mgat}, SLMRec\cite{tao2022self}, GRCN\cite{wei2020graph}, HHFAN\cite{cai2021heterogeneous}, MMGCL\cite{yi2022multi}, EgoGCN\cite{chen2022breaking}, A2BM2GL\cite{cai2022adaptive}, LUDP\cite{lei2022learning}       \\
\hline
\multirow{5}{*}{Textual} & \multirow{5}{*}{Reviews} & GRU          & VECF\cite{chen2019personalized}\\
                         &                          & Glove+MLP    & PAMD\cite{han2022modality}   \\
                         &                          & PV-DBOW    & JRL\cite{zhang2017joint}   \\
                         &                          & PV-DM    & MAML\cite{liu2019userdiverse}   \\
                         &                          & BERT    & DMRL\cite{liu2022disentangled}   \\
                         &                          & Word2Vec+SIF    & MKGAT\cite{sun2020multi}\\
\hline
\multirow{2}{*}{Textual} & \multirow{2}{*}{Title} & VILBERT        & MM-Rec\cite{wu2021mm}\\
                         &                          & Attention    & IMRec\cite{xun2021we}   \\
\hline
\multirow{5}{*}{Textual} & \multirow{5}{*}{Item description} &BERT         & PMGT\cite{liu2021pre}\\
                         &                          & TF-IDF    & ADDVAE\cite{tran2022aligning}   \\
                         &                          & Word2Vec+SIF    & MKGAT\cite{sun2020multi}\\
                         &                          & Sentence-transformer    & Lattice\cite{zhang_mining_2021}, FREEDOM~\cite{zhou2022tale}, HCGCN\cite{mu2022learning}, BM3\cite{zhou2022bootstrap}, DRAGON~\cite{zhou2023enhancing}\\
                         &                          & Sentence2Vec    & MVGAE\cite{yi2021multi}, EgoGCN\cite{chen2022breaking}, EliMRec\cite{liu2022elimrec}, InvRL\cite{du2022invariant}\\
\bottomrule
\end{tabular}
\caption{Modality feature extraction methods.}
\label{feature extraction}
\end{table}

\section{Model Classification}
We have classified models according to the methods that they utilized in the model as Table~\ref{tab: Model Classification} shows. The Matrix Factorization and MLP-based methods are the earlier ways to do recommendations. Previously, researchers just utilize the multimodality information simply, like VBPR~\cite{he2016vbpr} used the CNN extracted visual feature concatenate with id embeddings as the item representation then feeds into the MF module. Although it is simple, it could utilize the modality information as side information to help the cold start and data sparsity issues. Then a lot of deep learning techniques have been introduced into MMRec and show significant enhancement to utilize the modality information. For example, Autoencoder and Variational autoencoder techniques were introduced into the recommender system. Also, MMGCN~\cite{wei2019mmgcn} starts to use the Graph Convolutional Network to learn the representation of each modality then fuse the modality representation and id embedding together as the final item representations. Based on this GCN structure, researchers introduce knowledge graph~\cite{sun2020multi}, attention mechanism, edge refine~\cite{wei2020graph} and other fusion ways to help to learn better representations. In recent years, as the pretrain \& finetune has achieved good performance in CV and NLP, it is introduced into the MMRec system. This table only shows the overview of the classification of models, the following content will provide the introduction of each model and how they utilize the modality information by different techniques. One model might appear twice because some of the models are utilizing a combination of several approaches.

\begin{table}
\centering
\renewcommand\arraystretch{1.5}{
\begin{tabular}{m{19em}m{18em}}
\toprule
Method                 & Models                \\
\hline
Matrix Factorization & VBPR\cite{he2016vbpr}, GraphCAR\cite{xu2018graphcar}, VMCF\cite{park2017also}, AMR\cite{tang2019adversarial}, ACF\cite{chen2017attentive}, ConvMF\cite{kim2016convolutional}, DeepCoNN\cite{zheng2017joint}, DVBPR\cite{kang2017visually} \\
\hline
Multilayer Perceptron      & JRL\cite{zhang2017joint}   \\
\hline
(Variational) Autoencoder  & CKE\cite{zhang2016collaborative}, ADDVAE\cite{tran2022aligning}, MVGAE\cite{yi2021multi}\\
\hline
Attention Network    & UVCAN\cite{liu2019user}, ACF\cite{chen2017attentive}, MAML\cite{liu2019userdiverse}, MMRec\cite{wu2021mm}, DMRL\cite{liu2022disentangled}, IMRec\cite{xun2021we}  \\
\hline
Graph Neural Network(GNN)    & HHFAN\cite{cai2021heterogeneous}, MKGAT\cite{sun2020multi}, MMGCN\cite{wei2019mmgcn}, DualGNN\cite{wang2021dualgnn}, MGAT\cite{tao2020mgat}, MEGCF\cite{liumegcf}, GRCN\cite{wei2020graph}, Lattice\cite{zhang_mining_2021}, FREEDOM~\cite{zhou2022tale}, HUIGN\cite{wei2021hierarchical}, HGCL\cite{cai2022heterogeneous}, HCGCN\cite{mu2022learning}, EgoGCN\cite{chen2022breaking}, EliMRec\cite{liu2022elimrec}, InvRL\cite{du2022invariant}, A2BM2GL\cite{cai2022adaptive}, LUDP\cite{lei2022learning}, DRAGON~\cite{zhou2023enhancing}\\
\hline
RNN+Attention        & VECF\cite{chen2019personalized}\\
\hline
Matrix Factorization+CNN    & ConvMF\cite{kim2016convolutional}, DeepCoNN\cite{zheng2017joint}, DVBPR\cite{kang2017visually}\\
\hline
GNN+Pretrain and Finetune    & PMGT\cite{liu2021pre}, PAMD\cite{han2022modality}, MMCPR\cite{liu2022multi}\\
\hline
GNN+self-supervised learning    & SLMRec\cite{tao2022self}, BM3\cite{zhou2022bootstrap}, MMGCL\cite{yi2022multi}\\
\bottomrule
\end{tabular}}
\caption{Model classification}
\label{tab: Model Classification}
\end{table}

\subsection{Matrix Factorization based Models}
Matrix Factorization is a kind of Collaborative filtering method. The user-item interaction matrix is broken down into the product of two rectangular matrices with lower dimensions via the matrix factorization method.

Side information such as reviews and ratings has been used to help with recommendations. But the visual appearance of the items has been ignored before this work. VBPR~\cite{he2016vbpr} utilizes a pretrained deep CNN to get the visual feature then linearly transforms to visual space. Then the visual feature and id embedding are combined as the item representation and the representation is sent into Matrix Factorization to perform the preference prediction. AMR~\cite{tang2019adversarial} indicates the vulnerability of the current multimedia recommender system like VBPR and utilizes adversarial learning to build a more robust recommender system. VMCF~\cite{park2017also} constructs a product-affinity network with the nodes denoting products and the edges denoting the 'also-viewed' relations, which utilizes not only the appearance as side information but leverages the information hidden in the relations.

ConvMF~\cite{kim2016convolutional} would like to utilize the CNN architecture to capture the document information but the existing CNN method is used to solve the classification problem which is not suitable to do regression such as recommendation. This model then integrated probabilistic matrix factorization with CNN to do recommendations.

DVBPR~\cite{kang2017visually} incorporated the visual signals into the recommendation model, and it integrated the CNN module into the MF module that trains the image representation and recommendation model jointly to improve the performance. Additionally, by using this model, images that are similar to those in the training corpus could be produced. Given a user and category, this model with the Generative Adversarial Networks could generate a new item that is consistent with the user preference.

GraphCAR~\cite{xu2018graphcar} combines multimedia content with the traditional CF method. It first multiplies the user-item interaction matrix with the user attribute matrix or item feature matrix before using a two-layer graph convolutional network to generate the user and item latent representation. The preference score is then calculated using the inner product of the user and item latent vectors following a non-linear feature transformation. 

\subsection{Deep Learning based Models}
\subsubsection{Multilayer Perceptron.}\hfill 

A rich source of multimodal features could represent different aspects of user preference. In the JRL~\cite{zhang2017joint} framework, different sources of information have been projected into the unified space using MLP. Learning the weight to transfer multimodal features into  user and item embedding. Meanwhile, the associated user and item representations have been learned using the feature reconstruction loss.

\subsubsection{Convolutional Neural Network.}\hfill 

CNN is a kind of feedforward neural network with convolution layers and pooling operations to process the multimodal features. Most of the CNN-based recommender system utilizes CNN to extract features or learn the representation by capturing global and local features. 

Due to the bag of words model's inherent limitations, it's hard to fully capture the document information. ConvMF~\cite{kim2016convolutional} utilizes CNN to capture the local features of an image or document and bridge PMF and CNN to utilize both the interaction and description information.

DeepCoNN~\cite{zheng2017joint} utilizes the pre-trained word embedding technique to extract the semantic representation of both user and items from the review text. Then feed the user and item semantic embeddings into two parallel CNN modules to learn the item properties and user representations. The top layer is used to let the hidden factors of the user and items interacted with each other and jointly learn the item properties and user behaviors.

DVBPR~\cite{kang2017visually} Utilize the Siamese CNN framework and the last layer of CNN is used as the item representation. Those fashion-aware image representations are trained with the recommendation system together in an end-to-end manner. 

\subsubsection{Attention networks.}\hfill

The attention mechanism is motivated by human attention. The attention mechanism simulates that the user will pay different attention to different modalities or the different aspects of one modality, which could capture the user's preference and enrich the user representation.

The traditional collaborative filtering systems are not well designed for the multimodal recommendation that ignores leveraging the multimodal information. ACF~\cite{chen2017attentive} designs the item-level and component-level attention modules incorporating into traditional collaborative filtering models to do recommendations. The user's desire is typically represented by a fixed feature vector. However, different vectors for various items ought to better represent the user's preference. With a particular focus on diverse item characteristics, MAML~\cite{liu2019userdiverse} models user preferences for a variety of items. The retrieved text and visual features are first concatenated before being fed into the multilayer neural network. It introduces an attention neural network that utilizes the fusion of visual and text features, which can recognize users' varied preferences for various product attributes. DMRL~\cite{liu2022disentangled} also models the users' varied preferences but on the aspects of each modality. DMRL first utilizes a disentangled representation approach to learn the independent factors representation of several modalities. Based on this disentangled representation, it designs a weight-shared multimodal attention mechanism that could draw the user's attention to aspects of various modalities for factors. It may be possible to anticipate user preference for target items by combining such user and item representation with learned attention. DMR~\cite{wang2021multimodal} also learns the disentangled representations that could be capable to capture both complementary and common information from multi-modalities. Then it applies the disentangled representations to recommendation tasks.

The above micro-video recommendation works only consider the multimodal information of users but ignores the multimodal information of micro-video. UVCAN~\cite{liu2019user} learns the multimodal representation for both users and micro-video to do the personalized micro-video recommendation. unlike only concentrating on video attention, UVCAN utilizes the co-attention mechanism between items and micro-video to better jointly perform attention. A stacked attention network has been used to deal with the user profile and micro-video multimodal features, which considers the multimodal features as input queries and gets the video attention by multi-step reasoning. After learning the video representation, it will be leveraged as the input query to capture the user's attention through multi-step reasoning. 

News representation is critical for news recommendation, most of the existing news recommendation models ignore the visual information of news but only consider the text information like titles. MM-Rec~\cite{wu2021mm} incorporates both modalities to learn news representation. First, it does object detection to get the region of interest (ROIs) from the image. Then the visiolinguistic model has been used to encode both text and image ROIs and the co-attention transformers have been used to learn the inherent cross-modal relations. It also introduces a cross-modal candidate-aware attention network that learns the relations between clicked news and candidate news to better capture user preference on candidate news. 
The decision-making process is always ignored by researchers when they design the multimodal modeling module. IMRec~\cite{xun2021we} based on the fact that users pay more attention to the visual impression when they search for the news. It developed the local impression modeling module to extract significant clues from the impression image and bridged the gaps between semantics and impressions which enhance the news titles' semantic meaning. The global impression modeling module is used to fuse different cues without losing structural information. The experiment dataset makes use of the MIND dataset by adding screenshots of impression images.

\subsubsection{RNNs.}\hfill 

The existing fashion recommendation works mainly utilize multimodal information to model the news representation, but they did not consider that users will focus on different regions of the image to do decisions. VECF~\cite{chen2019personalized} uses the VGG model to get the pre-segmented image regions first. Then it models the human sense and utilizes an attention mechanism over the image regions to capture user preference. Also, the user reviews information has been used as a weak supervision signal to capture better preference of users by using Vanilla LSTM to infuse the attentive image embedding into the word generation. 
\subsubsection{Autoencoder.}\hfill

The collaborative filtering method is always hampered by the lack of user-item interactions. The rich side information has been used in the existing method to improve the recommendation system. CKE~\cite{zhang2016collaborative} investigates the way to utilize the collected heterogeneous information that has been gathered on the knowledge base. The network embedding procedure Bayesian TransR has been used to extract the structure knowledge embedding, Bayesian stacked denoising auto-encoder to extract the latent representation for text, and Bayesian stacked convolutional auto-encoder to extract the latent representation for the visual aspect. Then the collaborative joint learning procedure has been proposed. 

\subsubsection{Variational Autoencoder.}\hfill

Classical methods express the user preference as a single vector in the latent space which would not express the meaning facets. ADDVAE~\cite{tran2022aligning} utilizes the disentangled representations to capture better user preference that is influenced by several hidden factors. it leverages the text content to learn the second disentangled representation, coupling disentangled factors from two MacridVAE networks through mutual information maximization and then using the attention mechanism to align the representations with each other, which could resolve the sparsity of user-item interactions and map the uninterpretable dimensions from representations to words.

The sparsity of user-item interactions and feature noises will affect the graph embedding-based method employed for the recommendation. MVGAE~\cite{yi2021multi} is a multimodal variational graph auto-encoder model which utilizes the modality-specific variational encoder to learn the node representation. The Gaussian variable has been used for each node, the mean vector conveys semantic meaning and the variance vector represents the noise level of each modality. The unique modality attributes weight have been combined using the experts' product (PoE), which is dependent on the noise level.


\subsubsection{Graph Neural Networks} \hfill

\begin{figure}[H]
\centering
\subfigure[user-item bipartite graph]{
\includegraphics[width=6cm]{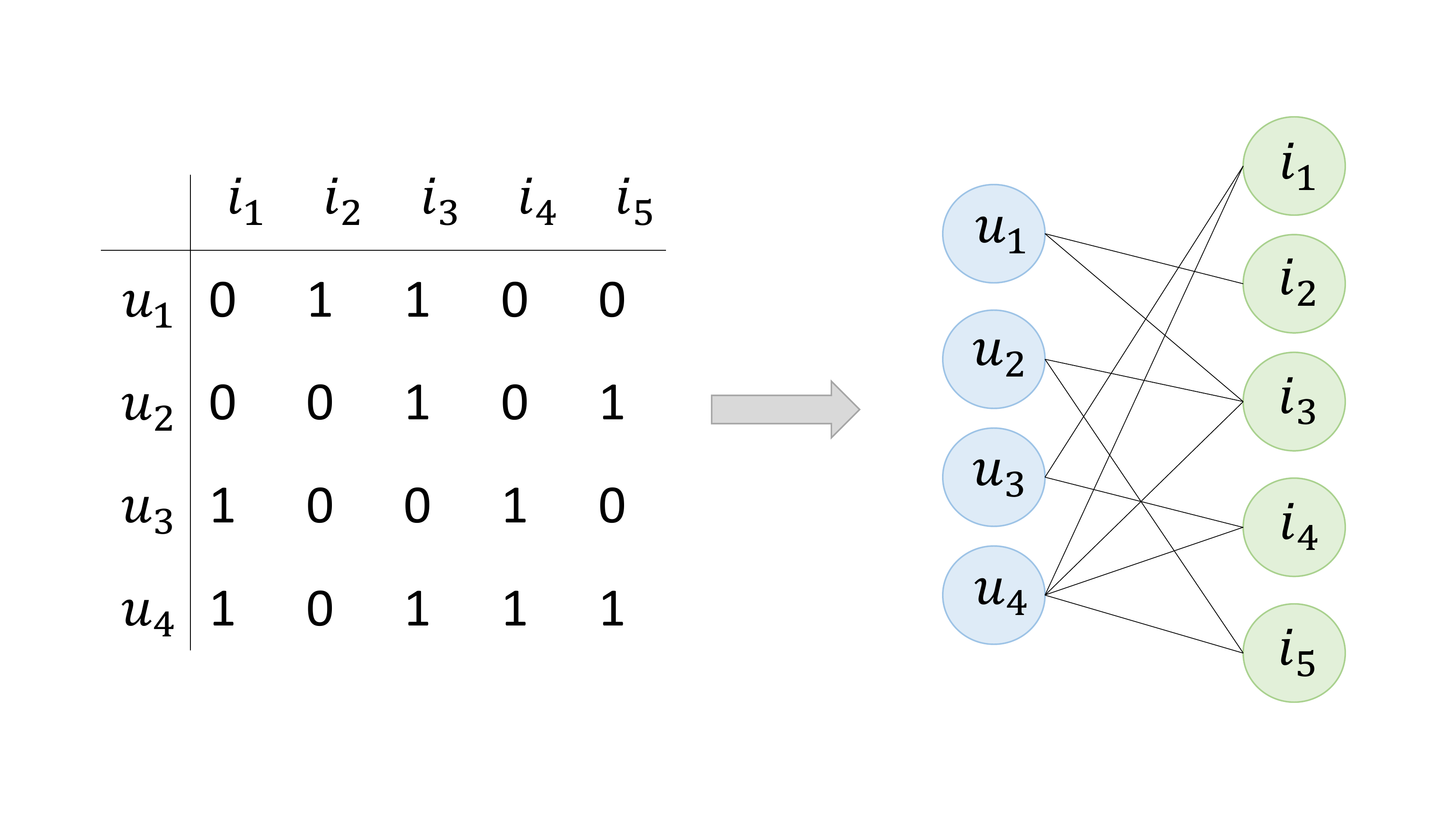}
\label{u-i}
}
\quad
\subfigure[user co-occurrence graph]{
\includegraphics[width=6cm]{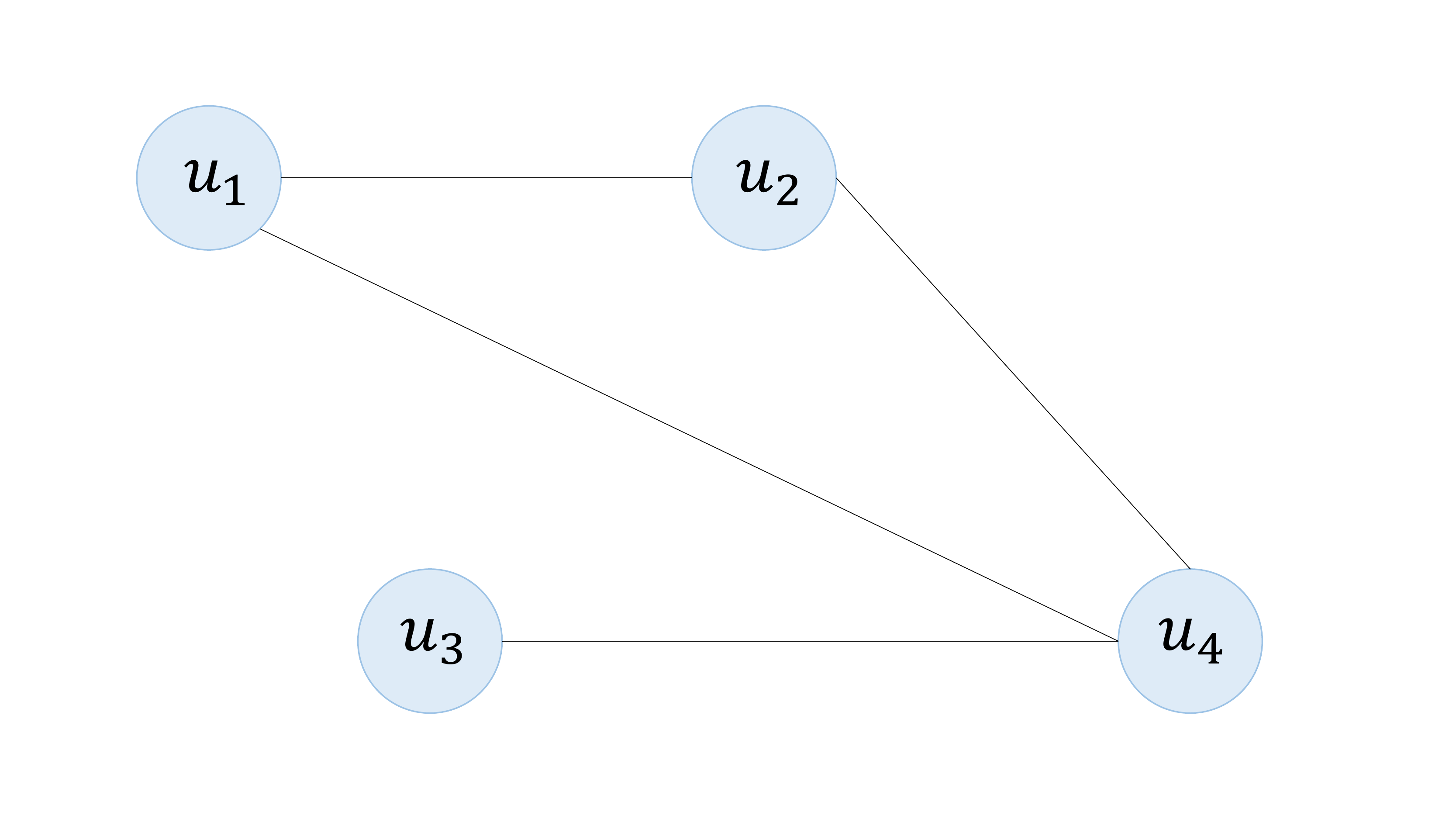}
\label{u-u}
}
\quad
\subfigure[item-item relation graph]{
\includegraphics[width=6cm]{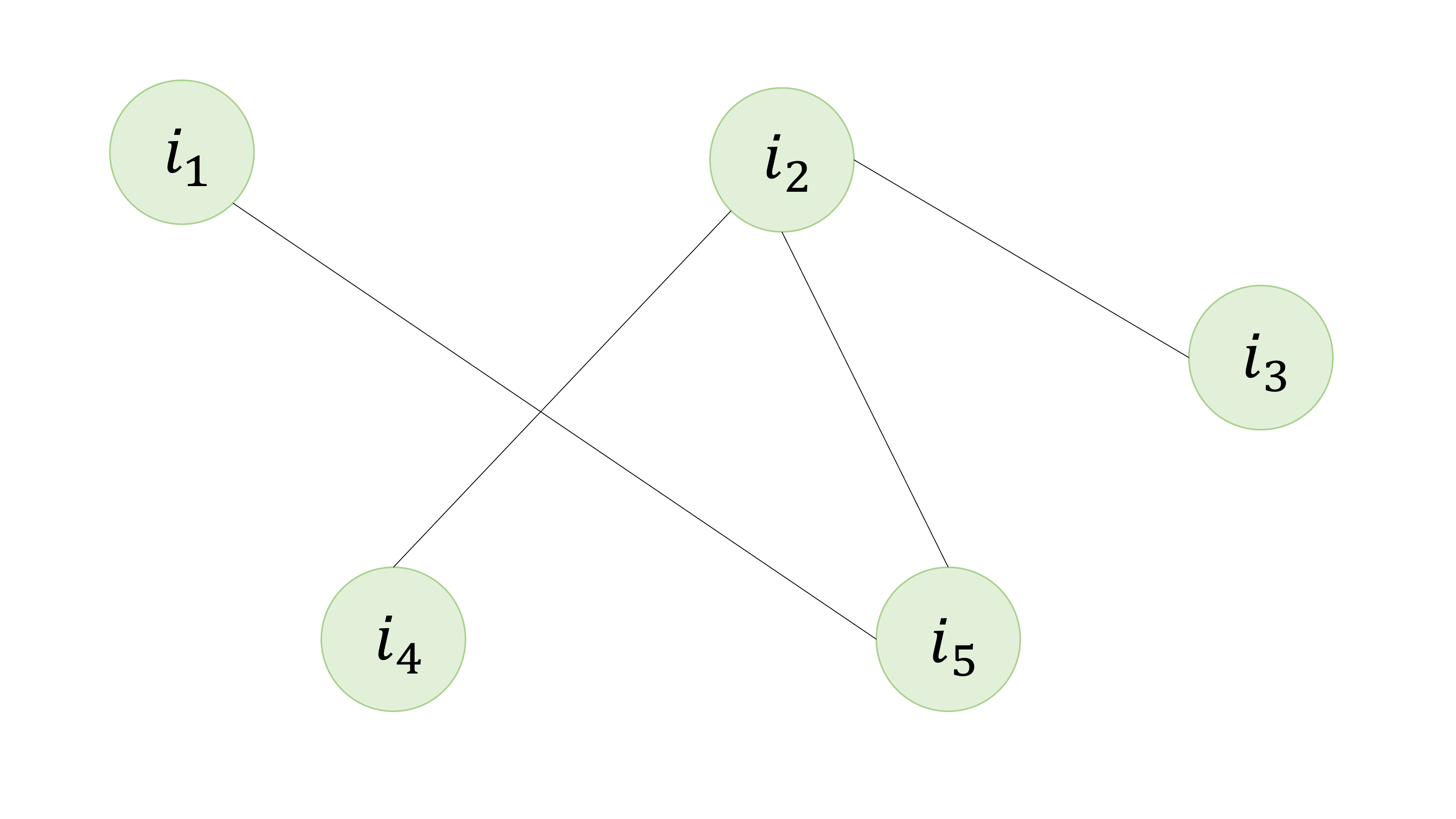}
\label{i-i}
}
\quad
\subfigure[knowledge graph]{
\includegraphics[width=6cm]{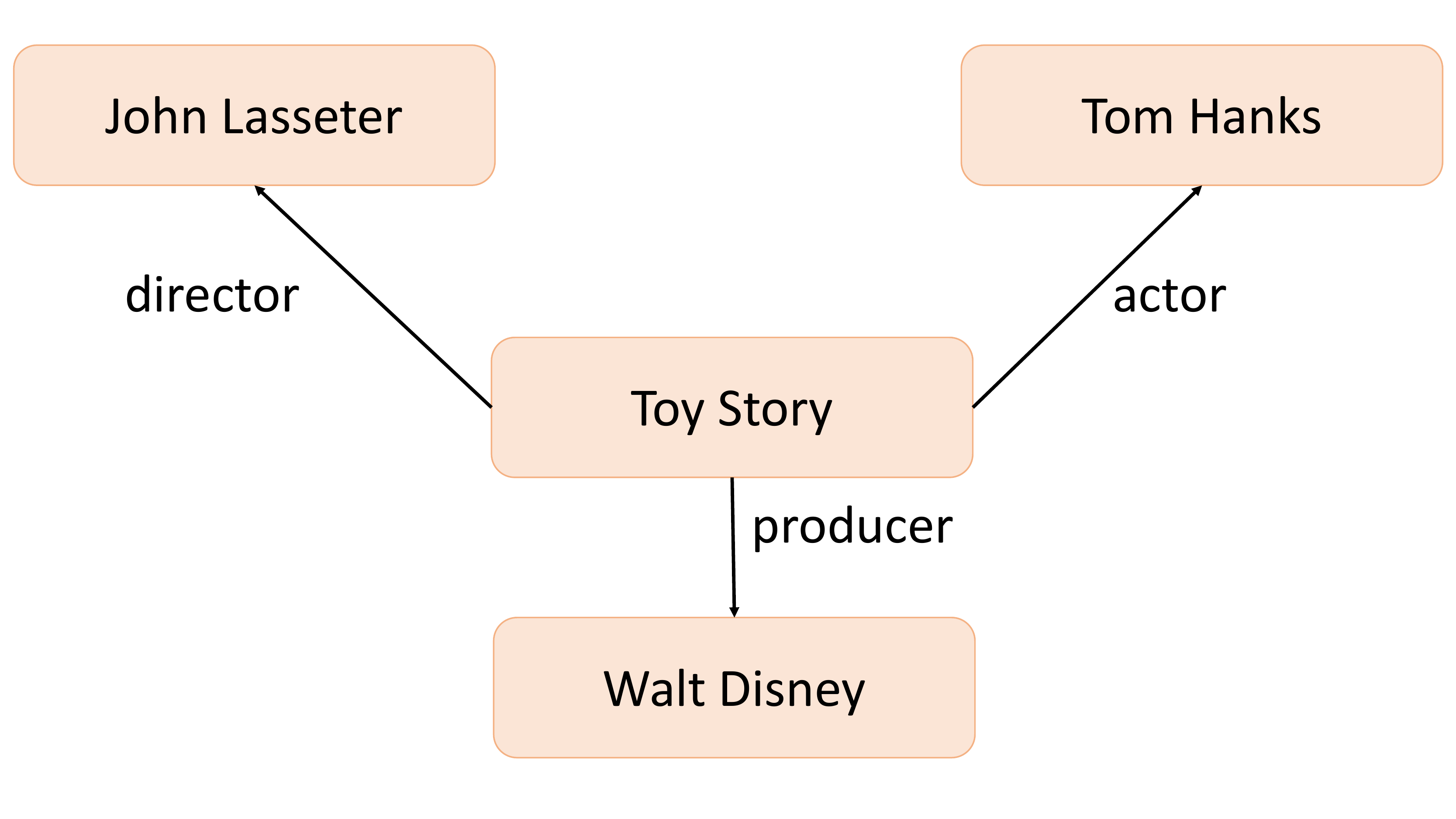}
\label{kg}
}
\caption{Graph structure in multimodal recommendation system}
\label{graph structure}
\end{figure}

Graph Neural Networks have been demonstrated powerful to learn the representations for graph data in different areas~\cite{guo2021syntax,zhou2020graph,ding2021graph,liu2022m2gcn,zhou2022inductive}. It could be easily extended to the recommendation system by considering the user-item relations as a graph relation. Most data in recommendation could be represented by the graph structure in Fig.~\ref{graph structure}. The main idea of GNN is to aggregate the information from neighbor nodes and update the node representation during the propagation process. We roughly divided those GNN-based models according to the fusion step.

\tb{Direct fusion.} Directly fuse multimodal embeddings extracted for each node. Use the multimodal feature as node representation then apply the Graph neural network techniques. Other models build the modality-specific graph to learn the representation for each modality, then fuse the output of GNN as the final representation. 

PMGT~\cite{liu2021pre} is the pre-trained model that leverages the fused multimodal feature and interactions. For the extracted multimodal embeddings, an attention mechanism has been used here to obtain the multi-model representation. Then the multimodal representation adds position embedding and role-based embedding together as the initialized node embeddings. Then the pre-train and downstream task starts based on the node embeddings.

The recommendation built on a knowledge graph as Fig.~\ref{kg} shows,  makes use of important outside knowledge to improve performance. However, they frequently disregard the different data types of multimodal features. MKGAT~\cite{sun2020multi} creates the multimodal knowledge graph using the entity-based approach. Different encoders are employed by the multimodal knowledge graph entity encoder to encode various data types, and the multimodal knowledge graph attention layer is used to aggregate neighbors' information. After each entity representation has been learned, the knowledge graph embedding will take them as input to learn the knowledge reasoning relation.

\begin{figure}[H]
    \centering
    \includegraphics[width=14cm]{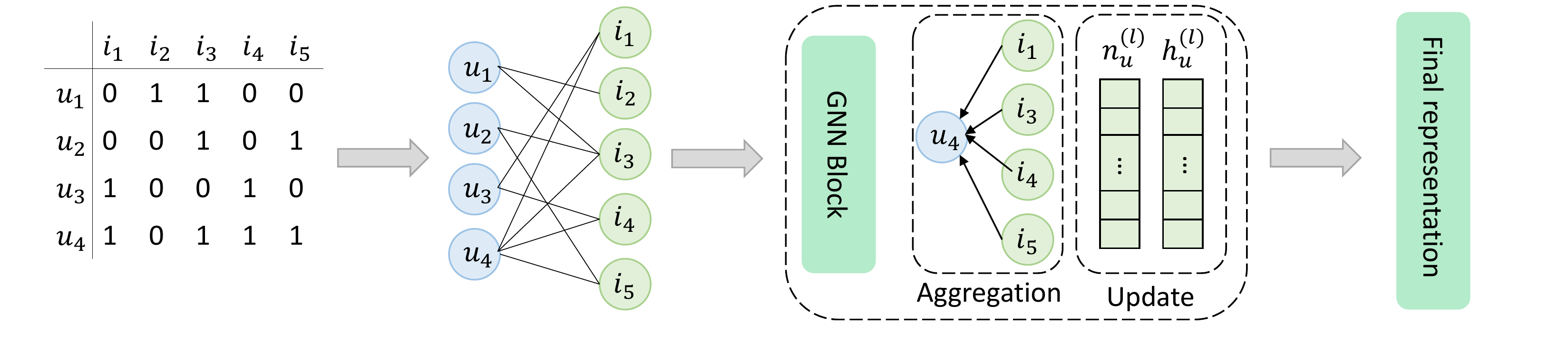}
    \caption{The overall framework of GNN in user-item bipartite graph}
    \label{fig: step}
\end{figure}

\tb{Heterogeneous graph fusion.} A heterogeneous graph is a kind of information network that contains different types of nodes. We consider the user and item as different types of nodes and the interactions between them as the edges as shown in Fig.~\ref{u-i}. Directly perform feature aggregation using GCN(MMGCN~\cite{wei2019mmgcn}, DualGNN~\cite{wang2021dualgnn}) or GAT(MGAT~\cite{tao2020mgat}) on the user-item bipartite graph as shown in Fig.~\ref{fig: step} for each modality. 

The previous multimodal recommendation works more focus on utilizing the multimodal information to represent items better but ignores the information interchanges between users and items. MMGCN~\cite{wei2019mmgcn} framework based on graph neural network to achieve message passing, and it builds the user-item bipartite graph for specific modalities. The information from the multi-hop neighbors could be used to better understand user preferences and create an improved user representation after the aggregate and combination processes have been completed. 

MMGCN simply applies GNNs for the interaction graph and equally treats the neighbors to do information change. It would be influenced by redundant even noisy information. MGAT~\cite{tao2020mgat} based on the MMGCN framework utilizes the original GCN to do aggregation and the same way to combine the aggregated result. To manage the information transmission for each modality, it added a new gated attention mechanism. As a result, it could record intricate user interaction patterns and produce better user recommendations. DualGNN~\cite{wang2021dualgnn} also introduces a model preference learning module and draws the user's attention to various modalities. Also, the influence made by the modality missing could be improved by the learned multimodal fusion pattern. 

The above GNN based-model like MMGCN merges isolated modal-specific graphs into a single one through operations such as concatenation, which is prone to underrate the interrelations among different modalities. EgoGCN~\cite{chen2022breaking} develops an effective graph fusion method Ego fusion to break the isolation. It constructed the modality-specific graphs the same as MMGCN but was not limited to propagating information in the unimodal graph by aggregating informative inter-modal messages of the neighbor nodes. The results of multimodal propagation and the ID embedding propagation are concatenated to be the final representations. 

Utilizing the non-uniform datasets with roughly fuse single modality feature to multimodal feature will lead to the single modal bias. EliMRec~\cite{liu2022elimrec} indicates that the single modal bias could be eliminated by leveraging causal inference and counterfactual analysis. The causal effect (Total Effect, TE) of single modality information on the final result is caused by (Total Indirect Effect, TIE) via single modal $\rightarrow$ item representation $\rightarrow$ final result and (Natural Direct Effect, NDE) via single modal $\rightarrow$ final result. The single modal bias can be eliminated by subtracting NDE from TE. It follows the same structure of MMGCN but eliminates the single-modal bias.

Multimodal feature extraction (MFE) and user interest modeling (UIM) are the two main modules of multimodal recommendation. But existing works ignore the mismatching between them. MEGCF~\cite{liumegcf} Utilizes the user sentiment information and semantic entities which are much related to user behavior and preference can help to change from the content-oriented MFE to user-oriented. To solve the mismatch problem, after transforming to user-oriented MFE, it makes the association between MFE and UIM using the Separated framework. Graph Convolutional Network as the commonly used technique that can capture the high order correlation between nodes is used here, but the sentiment information has been added to weight neighbor aggregation that could enhance the graph convolution operations. 

\tb{Homogeneous graph fusion.} A heterogeneous graph is a kind of information network that contains only one type of node and edge. The above GNN-based approach of micro-video recommendation only considers the relations between users and items but ignores the rich relations between the homogeneous data. Build homogeneous graphs by considering item-item relations(Lattice~\cite{zhang_mining_2021}) for each modality as Fig.~\ref{i-i} shows or considering the user co-occurrence frequency to build the graph as Fig.~\ref{u-u} shows, which helps to capture the relations between the same type of nodes.

The heterogeneous user-item interaction graph with multimodal information utilizes the historical interactions, but they ignore the semantic item-item relations which could better capture the item representation. LATTICE~\cite{zhang_mining_2021} utilizes the modality features in the modality-aware structure learning layer to learn item structures and then performs graph convolutions to learn better representation by the item affinities. Multimodal fusion introduces an attention mechanism that obtains the different importance of each modality and fuses the item embedding to get a powerful representation. Also, the self-supervised contrastive auxiliary task has been used to maximize the agreement between the item embedding of each modality and the fused multimodal representation. After learning the item representations, the existing collaborative filtering method will utilize them to get better performance. Not only the semantic item-item relations are used to build a homogeneous graph, but also the user co-occurrence time has been considered in DualGNN~\cite{wang2021dualgnn}. It builds on the user micro-video bipartite graph and user co-occurrence graph to draw the user's attention to various modalities. Also, the influence made by the modality missing could be improved by the learned multimodal fusion pattern. DRAGON~\cite{zhou2023enhancing} uses the user-item bipartite graph to learn the user-item relations. In addition, it utilizes item semantic graph and user co-occurrence graph to capture the internal relations among the two groups themselves. The dual representations for both users and items help to capture complementary information.

There is no recommendation system that considers the user's intent before. In order to direct the learning of user and item representations, HUIGN~\cite{wei2021hierarchical} tries to learn the multi-level user intents. It constructs two different types of information aggregation methods and a co-interacted item graph. The intra-level aggregation leverages the correlation between items and gets the representations that could convey the user intent to items. The inter-level aggregation models the relations between high-level and low-level user intents. The supernodes were established as coarser-grained user intents.

HGCL~\cite{cai2022heterogeneous} propose that most of the GNN techniques (\eg GCN~\cite{kipf2016semi}, GraphSage~\cite{hamilton2017inductive}, GAT~\cite{velivckovic2017graph}) are designed for the homogeneous graph but users and items in the recommendation system are the heterogeneous data. The performance of the present GNN-based recommendation systems will be impacted by their inability to accurately capture the heterogeneity of nodes in the user-item graph and their over-smoothing issue. In HGCL~\cite{cai2022heterogeneous}, the heterogeneous graph encoder network is used to acquire the heterogeneity in the user-item bipartite networks by utilizing the random surfing model to generate the specific node-type homogeneous graphs, then apply a GNN on the homogeneous graphs to learn the representations. Also, a graph contrastive learning framework also applies to the graph with specific node types by averaging the mutual information between global representation and local patch representation to learn an expressive representation of nodes. 

The commonly used modality fusion method is to fuse the multimodal information into the id embedding to learn meaningful item representations. LATTICE begins to figure out that modality features could be used to construct the latent structure between items but not fuse the modality information into the user-item bipartite graphs. FREEDOM~\cite{zhou2022tale} finds that learning latent structure is inefficient and freezes the item-item graph before training could help to learn better item semantic relations. It also denoises the graph based on the calculated scores of the degree-sensitive edge pruning technique. HCGCN~\cite{mu2022learning} also follows the general process of LATTICE by constructing the item semantic graph and user-item interaction graph. It additionally introduces the co-clustering technique that learns the user and item representations in the same subgraph, which helps to capture the various user behavior patterns. The corresponding co-clustering and item-clustering losses have been used to enhance the user-item preference feedback and adjust the modality importance. A similar structure that utilizes features to construct a homogeneous graph is also mentioned in A2BM2GL~\cite{cai2022adaptive}. It leverages the similarity of multimodal contents to construct the feature graph to learn the semantic representation. Except for the semantic representation learning module, it constructs a collaborative representation learning module to learn the collaborative representation and an anti-bottleneck module that utilizes an attention mechanism to learn the importance weights of the short-range and long-range nodes. It updated the weighted regression loss to adaptive recommendation loss that could help to capture the varying intensity of users. LUDP~\cite{lei2022learning} indicates that the user preference is poorly learning using the current models that only enrich the item representation by multimodal features. So, it designed both the same item-item modal similarity graph to capture the potential item relations and the user preference graph to capture the user preference on different modalities.

\tb{Sampled subgraph fusion.} Sample a subgraph for each node. Group nodes based on modalities. HHFAN~\cite{cai2021heterogeneous} develops a heterogeneous graph that contains user, item and multimodal information. A random walk through the graph to sample neighbors according to the node type. It employs the FC layer to transform various types of node vectors into a single space and LSTM to aggregate embeddings of nodes of the same type for each set of nodes in the intra-type feature aggregation network. Then uses attribute-aware self-attention and neighbor-aware attention for different types of nodes in the inter-type feature aggregation network to create final representations that are better able to convey the structure and semantic information. HCGCN~\cite{mu2022learning} sample subgraph for the user-item interaction graph to divide users and items into groups, which helps to capture the various user behavior patterns for each group. HCGCN samples subgraph to capture the user behavior patterns, a similar method has been utilized by InvRL~\cite{du2022invariant}. It indicates the spurious correlation issue that the generic representation focus on the semantic meaning but not the factors that influence user preference. To alleviate the impact of spurious correlation, it sampled subgraphs to learn the invariant masks which divided the multimodal representation into invariant and variant representations. Based on the UltraGCN~\cite{mao2021ultragcn}, utilizing the invariant representations could eliminate the spurious correlation. 

\tb{Use multimodal info to refine the graph structure.} The recommendation system extensively uses the user-item interaction graph to perform representation learning and information propagation. However, there are some false-positive edges exist which will influence the performance of the recommendation. GRCN~\cite{wei2020graph} works on identifying and cutting the noise edges to refine the structures of the user-item interaction graph. In the graph refine layer, a neighbor routing mechanism has been used to refine the prototypical network. The matching between user preference and the prototypical shows the probability to be the true-positive edge. According to the matching score, those edges will be corrupted. Then the graph convolutional layer and prediction layer will be used to do recommendations based on the refined user-item graph. LATTICE~\cite{zhang_mining_2021} builds a graph structure learning layer with modality-aware to learn item graph structures from multimodal information and combine multimodal graphs. By using graph convolutions, one item can acquire useful high-order affinities from its neighbors along the learned graph structures.

\subsubsection{Self-supervised learning}\hfill 

The multimodal recommendation aims to learn suitable representations for each modality. They are mainly based on historical interactions and the multimodal side information to do supervised learning. The self-supervised learning techniques could also help to explore the underlying relations. SLMRec~\cite{tao2022self} learns the relationships by incorporating self-supervised learning into GNN-based recommendation models. This paper treats ID as a special modality with visual, textual and acoustic modalities. And it constructs the modality-specific graph the same as ~\cite{wei2019mmgcn,wang2021dualgnn,zhang_mining_2021}. Then the GNN techniques are used in the modality-specific graphs and get the modality representations. It considers each modality as a separate feature and utilizes the self-supervised learning tasks feature drop(FD), feature masking (FM) and feature fine and coarse spaces (FAC) to produce different views of items then uses contrastive learning to learn the underlying training signals. 

BM3~\cite{zhou2022bootstrap} overcomes the computational cost and noise supervision signal issues by introducing self-supervised learning methods into the multimodal recommendation. First, it utilizes simple dropout techniques to generate different views from the item and user representations. Then three multimodal objectives were optimized to find better representations. Graph Reconstruction Loss is used to maximize the prediction of the item when giving the user by reconstructing the user-item interaction graph. The Inter-modality Feature Alignment Loss and the Intra-modality Feature Masked Loss were both used to align modality features. This model achieves better performance and also consumes fewer computational resources. 

MMGCL~\cite{yi2022multi} introduces the self-supervised learning approaches based on graphs into the micro-video recommendation and it introduces a cutting-edge negative sampling method that demonstrates the relationship between modalities. Two data-augmented techniques have been used in this model: modality edge dropout and modality masking. Another technique used is to perturb one of the negative sample's modalities making just one modality difference between positive and negative samples. These methods combine to reveal the relationships between several modalities and ensure that each modality makes a meaningful contribution.

\subsubsection{Pretrain}\hfill 

The pre-trained models have been proven useful in natural language and images. Side information like images and text has been used to enhance the effectiveness of the recommendation system. PMGT~\cite{liu2021pre} develops a pre-trained model which considers both the multimodal features and item relations to learn the item representation. An attention mechanism has been used to aggregate the multimodal features by weight and a graph structure has been constructed by the common user behaviors to represent the item relations. The two pre-train tasks are graph structure reconstruct and masked node feature reconstruction. It developed a mini-batch contextual neighbors sampling algorithm to sample negative and positive data which is used to handle large-scale graphs. After the Pre-train has been done, we could get the node representations and use them to initiate the item embeddings for the downstream tasks. In this paper, the recommendation task uses Neural Collaborative Filtering as the fundamental model. Training NCF and fine-tuning the item embeddings depending on the user-item interactions achieved the best accuracy compared with the other multimodal recommendation models.

The homogeneous graphs like the item graph and the user graph are proven helpful to learn the representation and improve the performance of recommendations. MMCPR~\cite{liu2022multi} pretrains on the homogeneous and heterogeneous graphs, aiming to learn better user and item representations that could capture user preference and item characteristics. To learn user and item representation, the user graph and item graph are built based on their co-interact relationships. For user representation, it considers two modalities: review text and user graph. 
It uses LightGCN as the graph encoder to get the node representations in the user graph. Intra-modal aggregation is used for the texts of the reviews to get the aggregated review representations and inter-modal aggregation aggregates the two representations of modalities by attention mechanism. For item representation, it considers three modalities: text, image and item graph. The CLIP is used to extract the features. Because the image and text are describing the same item but from different aspects, the self-supervised contrastive inter-modal alignment (CIMA) task is used to align these two embeddings by maximizing the cos similarities between them. The intra-modal aggregation is the same as the user side to aggregate those three representations by attention mechanism to get the final item embeddings. Pre-train based on the Contrastive Inter-modal Alignment loss and user-item similarity loss to get the pre-trained embeddings. Then the LightGCN based on MixGCF is selected as the fine-tuning base model.  

The multimodal features reflect both common knowledge and specific knowledge of the items. Recent simple fusion techniques combine both common and unique characteristics into a single vector representation. PAMD~\cite{han2022modality} seeks to understand both common and unique characteristics of objects. The modality characteristics are divided into common and particular representations using a disentangled encoder. Then contrastive learning performs cross-modality alignment to map the modality-common representations with primitive representations. Then it utilizes the transferred representations of primitive and specific-modality representations into optimization by measuring the gaps between them. After the pre-training stage, every item could get four representations. It uses the attention mechanism to fuse all the embeddings to get the final representations and adopts BPR as the downstream base model.

\begin{table}
\centering
\renewcommand\arraystretch{1.5}{
\begin{tabular}{m{9em}m{9em}m{19em}}
\toprule
Fusion Step &  Fusion Method               & Models                \\
\hline
\multirow{2}{*}{Early Fusion} & Concatenate  & VBPR\cite{he2016vbpr}, UVCAN\cite{liu2019user}, MAML\cite{liu2019userdiverse}\\
& Attention-based  & VECF\cite{chen2019personalized}, PMGT\cite{liu2021pre}, LATTICE\cite{zhang_mining_2021}, FREEDOM~\cite{zhou2022tale}, HCGCN\cite{mu2022learning}, InvRL\cite{du2022invariant}, LUDP\cite{lei2022learning} \\
\hline
\hline
\multirow{4}{*}{Intermediate Fusion} &  Concatenate & MMGCL\cite{yi2022multi} \\
& Attention-based  & MMRec\cite{wu2021mm}, MKGAT\cite{sun2020multi}, DMRL\cite{liu2022disentangled}  \\
& Product of Expert & MVGAE\cite{yi2021multi}\\
& Ego Fusion & EgoGCN\cite{chen2022breaking}\\
\hline
\hline
\multirow{3}{*}{Late Fusion} &   Concatenation   & MMGCN\cite{wei2019mmgcn}, GRCN\cite{wei2020graph}, SLMRec\cite{tao2022self}, JRL\cite{zhang2017joint}, EliMRec\cite{liu2022elimrec}   \\
&   Element-wise sum   & MMGCN\cite{wei2019mmgcn}, SLMRec\cite{tao2022self}, A2BM2GL\cite{cai2022adaptive}   \\
&   Attention-based   &  DualGNN\cite{wang2021dualgnn}, MGAT\cite{tao2020mgat}, PAMD\cite{han2022modality}, MMCPR\cite{liu2022multi}, DRAGON~\cite{zhou2023enhancing}\\
\bottomrule
\end{tabular}}
\caption{Modality fusion}
\label{tab: fusion}
\end{table}

\section{Modality fusion}

For the multimodal recommendation system, modality fusion is an important research point. Finding an explainable and complementary fused multimodal representation can achieve significant enhancement. In technical terms, multimodal fusion is integrating the information from different modalities to get a single representation that could be used in different tasks like prediction and classification. The advantages of multimodal fusion are three points. Firstly, utilizing multimodal information could capture the universal pattern in each modality which could get a more robust representation based on the same pattern. Secondly, the multi-modality can capture complementary information which is not visible in the single modality. Lastly, the multi-modality fusion could handle the modality missing problem that could still operate with one modality missing. According to the timing of action in the pipeline, it could be roughly divided into early fusion, late fusion and intermediate fusion~\cite{baltruvsaitis2018multimodal}.
\begin{itemize}
\item Early fusion also names data-level fusion which means it fuses modality embeddings into a single feature representation and then inputs to the model. But this approach could not capture the complementary information of multi-modality and might include data redundancy. The early fusion approach is usually applied with some feature extraction methods (like PCA and autoencoder) to overcome the problem. 
\item Late fusion also named decision-level fusion based on the prediction result or scores of each modality. The commonly used approaches are max-fusion, averaged-fusion, Bayes’rule based and ensemble learning. This kind of approach could learn complementary information and each modality is independent from others which will not lead to the addition of mistakes.
\item The intermediate fusion fuse the modality information after getting the high dimensional embeddings of each modality and then utilize the middle layer to do fusion. This kind of fusion depends on the model design.
\end{itemize}

According to the different fusion methods, we can divide it into element-wise sum, attention-based and concatenation. The element-wise sum and concatenation simply add or concatenate the modality representation, formally:
\begin{align}
    \bm{u}&=\bm{u_v}+\bm{u_t}+\bm{u_a} & \bm{u}&=\bm{u_v}\parallel \bm{u_t}\parallel \bm{u_a}
    \label{eq1}
\end{align}

The attention-based method is used to capture the different importance of each modality, and it could be applied to the basic fusion method like the element-wise sum:
\begin{equation}
    \bm{u}=\sum_{m \in \left \{ v,t,a \right \} }^{} \alpha_m\bm{u_m}
    \label{eq2}
\end{equation}

There exists some model designed their own way to fuse modalities like MVGAE~\cite{yi2021multi} adapt the Product of Expert module to distinguish modalities with multiple uncertainty levels that solve the issue of concatenation. And IMRec~\cite{xun2021we} does not fuse the modality information but encoded the impression image as a whole. We classify the models according to their modality fusion method as Table~\ref{tab: fusion} shows.

\section{Measurement and Optimization}
\subsection{Evaluation metrics}
Commonly used evaluation metrics for the recommendation system include accuracy, precision, recall, F1-score, hit rate, Normalized Discounted Cumulative Gain (NDCG) and Mean Average Precision (MAP). We consider the ground truth labels as positive and negative, within the predicted labels, there will be True Positive (TP), True Negative (TN), False positive (FP) and False Negative (FN). Accuracy represents the ratio of corrected predicted items over the total number of items:
\begin{equation}
    Accuracy = \frac{TP+TN}{TP+TN+FP+FN} 
    \label{eq3}
\end{equation}

In a real situation, the positive and negative samples in datasets are unbalanced. Usually, the positive samples are less than 10\%, which means you can get an accuracy higher than 0.9 even if you predict all items as negative. We aim to measure the ability to predict positive items. Recall and Precision are used to represent the ratio of the truly predicted positive samples over the predicted positive and the ground truth positive. F1-score combines the result of recall and precision:
\begin{align}
    Recall &= \frac{TP}{TP+FN} &  
    Precision &= \frac{TP}{TP+FP} &
    F1-score &= \frac{2*precision*recall}{precision+recall}
    \label{eq4}
\end{align}

The hit rate likes the name implies, represents the ratio of hit items in the predicted top-k list over the total items in the test set. If a user actually interacts with one of the top-k items we recommend, it is considered a hit.
\begin{equation}
    HR = \frac{NumberOfHits@K}{Test} 
    \label{eq5}
\end{equation}

Precision and Recall do not consider ordering. The average precision (AP) is used to evaluate order-sensitive recommendation results for one user, and MAP is the average of $AP@K$ over all users:
\begin{align}
    AP@K &= \frac{1}{\left | K \right | }\sum_{k=1}^{K} \frac{p_{k}\cdot rel(k)}{k}  & MAP&=\frac{1}{\left | U \right | } \sum_{u \in U}^{} AP_u
    \label{eq6}
\end{align}

Normalized Discounted Cumulative Gain (NDCG) is also a measure of rank quality. Cumulative Gain (CG) is the relevance score in a recommendation list, DCG fills the gaps that CG only utilizes the relevance score but did not consider the rank position. iDCG is the DCG of the ideal ordered recommendation list. NDCG is calculated by DCG over iDCG, formally:
\begin{align}
    DCG&=\sum_{i=1}^{K} \frac{2^{relevance_i}-1 }{log_2(i+1)}  & NDCG=\frac{DCG}{iDCG}
    \label{eq7}
\end{align}

\begin{table}
\centering
\renewcommand\arraystretch{1.5}{
\begin{tabular}{m{9em}m{25em}}
\toprule
Method               & Models                \\
\hline
BPR  & VBPR\cite{he2016vbpr}, ACF\cite{chen2017attentive}, MMGCN\cite{wei2019mmgcn}, DualGNN\cite{wang2021dualgnn}, GRCN\cite{wei2020graph}, MGAT\cite{tao2020mgat}, GraphCAR\cite{xu2018graphcar}, PAMD\cite{han2022modality}, MMGCL\cite{yi2022multi}, MKGAT\cite{sun2020multi}, MVGAE\cite{yi2021multi}, DVBPR\cite{kang2017visually}, LATTICE\cite{zhang_mining_2021}, FREEDOM~\cite{zhou2022tale}, DMRL\cite{liu2022disentangled}, HCGCN\cite{mu2022learning}, EgoGCN\cite{chen2022breaking}, EliMRec\cite{liu2022elimrec}, InvRL\cite{du2022invariant}, LUDP\cite{lei2022learning}, DRAGON~\cite{zhou2023enhancing} \\
\hline
Pairwise Loss & MAML\cite{liu2019userdiverse}\\
\hline
MSE Loss & VMCF\cite{park2017also}\\
\hline
Cross Entropy Loss & UVCAN\cite{liu2019user}, MMRec\cite{wu2021mm}, MMCPR\cite{liu2022multi}, IMRec\cite{xun2021we}\\
\hline
Contrastive Loss & PAMD\cite{han2022modality}, MMCPR\cite{liu2022multi}. MMGCL\cite{yi2022multi}, BM3\cite{zhou2022bootstrap}, SLMRec\cite{tao2022self}\\
\hline
Reconstruction Loss & PMGT\cite{liu2021pre}, BM3\cite{zhou2022bootstrap} \\
\hline
Self-designed Loss&A2BM2GL\cite{cai2022adaptive}\\
\bottomrule
\end{tabular}}
\caption{Loss Function}
\label{tab: loss}
\end{table}

\subsection{Objective Functions}
Typical losses include mean square loss (MSE), pairwise loss, cross-entropy loss, and Bayesian Personalized Ranking (BPR) loss. When it comes to the top-k suggestion, the final two loss functions are typically utilized but MSE and pairwise loss are used less. MSE is always used for rating prediction. 
\begin{equation}
    MSE=\sqrt{\frac{1}{m}\sum_{i=1}^{m}(h(\bm{x^i})-\bm{y^i})^2} \\
    \label{eq8}
\end{equation}

Basic pairwise loss is not usually used but some updated version exists. The Weighted ApproximateRank Pairwise (WARP)~\cite{weston2010large} updated the basic pairwise loss by considering the ranking that penalizes a positive item at a lower rank much more heavily than the one at the top.
While BPR~\cite{rendle2012bpr} approaches the top-k task as a ranking problem, encouraging the ranking of positive things above negative ones for the given user, Cross entropy loss approaches the top-k task as a classification challenge and make the target user u and the target item i as similar as possible. BPR loss is more suitable for the top-k recommendation and most of the models are applying this loss to do optimization like  ~\cite{wei2019mmgcn,wang2021dualgnn,zhang_mining_2021,chen2019personalized}.
\begin{align}
    \mathcal{L}_{bpr}&=\sum_{(u,i,i')\in \mathcal{R}}^{}  -ln\mu (\bm{u}^T\bm{i}-\bm{u}^T\bm{i'}) \\ \mathcal{L}_{CE}&=-\sum_{(u,i)\in \mathcal{R}}^{}y_{ui}\cdot log(\bm{u}^T\bm{i})+(1-y_{ui})\cdot log(1-\bm{u}^T\bm{i})
\end{align}

As the pretrain and finetune were introduced into Multimodal recommendation, the pretrain task loss has been used to get a better representation result. \cite{liu2021pre} utilizes the edge reconstruction loss and the node feature reconstruction loss to do optimization. Recently, contrastive loss has also been introduced into recommendation as the self-supervised learning method has been applied in this area. Contrastive loss aims to increase the similarity between similar items but make the item representation different from the other items. And the InfoNCE loss is the variant that considers the problem as a multi-class classification~\cite{tao2022self, yi2022multi}. BM3~\cite{zhou2022bootstrap} doesn't need to do negative sampling and designs different views to get the contrastive loss by calculating the similarities between views.
\begin{equation}
    \mathcal{L}_{CL}=\frac{\bm{u}^T\bm{i}}{\left \| \bm{u} \right \|_2\left \| \bm{i} \right \|_2  } 
\end{equation}

SLMRec~\cite{tao2022self} has designed the contrastive loss for both self-supervised tasks and the recommendation task.
\begin{equation}
    \mathcal{L}_{InfoNCE}=-log\frac{exp(\bm{u}^T\bm{i}/\tau )}{ {\textstyle \sum_{i' \in N_u}^{}exp(\bm{u}^T\bm{i'}/\tau)} }
\end{equation}

We have classified the models according to the objective function they used as Table~\ref{tab: loss} shows.

\section{Datasets and Experiments}
\subsection{Experimental Datasets}
Some commonly used datasets are huge with a large number of users, items and interactions like, Kwai and Tiktok, and some of the datasets are not available to the public. Here we use the publicly available and most commonly used Amazon dataset and one of the Food datasets collected from a website as our experimental dataset. We conduct the experiment on the four main categories baby, sports, electronics and food. The statistics of these four datasets are summarized in Table~\ref{tab: dataset}. 
Same as the previous works~\cite{wei2020graph,wei2019mmgcn,zhang_mining_2021,zhou2022bootstrap}, we do the 5-core setting for all the amazon datasets, the 10-core setting for the food dataset and split them into training, validation, and testing sets with a ratio of 8:1:1. It should be mentioned that the split is based on randomly selecting train and test samples for each user according to the ratio. In the following experiments, we will discuss different ways to split train, test and validation sets such as based on the user time or global time. 

The metadata of the amazon datasets contains the text description and the URL of images that could be used as textual and visual modalities. For the collected Food dataset, the recipe and ingredients are used as the text modality, the food images are the visual modality. We used the pre-trained sentence-transformers~\cite{reimers2019sentence} to extract the text features with 384 dimensions and follows~\cite{zhang_mining_2021} to use the published 4096-dimensional visual features. After getting the textual and visual embeddings, we consider the user ratings as a positive interaction and then construct the user-item graph based on the review ratings. 

\begin{table}
\caption{Statistics of the datasets.}
\small
\centering
\begin{tabular}{ l |c c c |r } 
\hline
Dataset & \# Users & \# Items & \# Interactions & Sparsity \\
\hline
Baby & 19,445 & 7,050 & 160,792 & 99.8827\% \\
sports & 35,598 & 18,357 & 296,337 & 99.9547\% \\
FoodRec & 61,668 & 21,874 & 1,654,456 & 99.8774\%\\
Elec &  192,403& 63,001 & 1,689,188 & 99.9861\% \\
\hline
\end{tabular}
\label{tab: dataset}
\end{table}

\subsection{Baseline models}
We provide a framework for the readers to run their datasets using baseline models or implement their own models based on the framework. The framework is implemented by Pytorch and the input format is simple with textural embedding, visual embedding and the interaction graph. The following models except ADDVAE~\cite{tran2022aligning} are already implemented in our framework and the experiments are conducted to compare their recommendation performance. note that BPR and LightGCN are not using multimodal information but only use historical user-item interactions. The ADDVAE model is not implemented in our framework because it utilizes the text feature in a different input format produced by the TF-IDF extraction method. We have implemented ADDVAE separately and run it with the same train-test splitting sets and text raw feature with different extraction methods. Here we just include them as the baselines of multimodal recommendation performance.
\begin{itemize}
    \item BPR~\cite{rendle2012bpr}
    utilizes the historical user-item interactions to model user and item representations as the latent factor. Predicting user's preference based on the similarity between the representations. This model introduces the mainly used loss function BPR loss.
    \item LightGCN~\cite{he2020lightgcn}
    is a graph-based model designed by simplifying the Graph Convolution Network (GCN)~\cite{kipf2016semi} which improves the performance and also the efficiency of recommendation.
    \item VBPR~\cite{he2016vbpr}
    is the first model that considers the visual features in the recommendation. It integrated the visual embedding into item representations. To be fair, we concatenate both vision and text features as the multimodal feature when we learn the representations. The BPR loss is used here to learn the user preference.
    \item MMGCN~\cite{wei2019mmgcn}
    learns the modality-specific preference by constructing the modal-specific graphs for each modality. And utilize those user-item graphs to learn the representations in each modality. Then aggregates all the modality-specific representations to get the final user and item embeddings.
    \item DuaGNN~\cite{wang2021dualgnn}
    introduces a new user-user co-occurrence graph that utilizes the representations learned from modality-specific graphs and fuses the neighbor's representations.
    \item GRCN~\cite{wei2020graph}
    utilizes the same user-item graphs as previous GCN-based models generated. It identifies the false-positive edges in the graph and cut off the edges then the new representations would be generated by performing aggregation and information propagation in the refined graph. 
    \item LATTICE~\cite{zhang_mining_2021}
    introduces a new item-item graph for each modality and gets the latent item graphs by aggregating all the modalities. Both the item-item graph and user-item graph utilize the graph convolution operations to get out the final user and item representations.
    \item BM3~\cite{zhou2022bootstrap}
    introduces a new self-supervised learning method to help reduce the cost of computational resources and the issue of wrong supervision signals. It utilizes the simple dropout technique to generate contrastive views and use three contrastive loss functions to optimize the result representations.
    \item SLMRec~\cite{tao2022self}
    utilizes the self-supervised learning techniques in the Graph Neural network based model to learn the underlying relations. The SSL tasks supplement the supervised tasks to uncover the hidden signals from the data itself. The FD, FM and FAC self-supervised data augmentation operations have been performed here and then the contrastive loss is used to optimize.
    \item ADDVAE~\cite{tran2022aligning}
    utilizes the MacridVAE network with mutual information maximization to get two disentangled representations from user-item interactions and the text content of users. Then utilizing the attention mechanism to align the disentangled representations. 
    \item FREEDOM~\cite{zhou2022tale}
    utilizes the item-item graph for each modality established the same as LATTICE but freezes the graphs before training, and introduces the degree-sensitive edge pruning techniques to denoise the user-item interaction graph.
\end{itemize}

\subsection{Experimental Results}
In this section, we first analyze the performances of all implemented algorithms on the four datasets (Section 7.3.1). We then discuss the performance of different types of data split strategies (Section 7.3.2). Finally, we evaluate the effectiveness of multimodal representation learning by utilizing different modality information(Section 7.3.3). We adopt the widely-used metrics to evaluate the performance of preference ranking: Recall@k and NDCG@k. By default, we set k = 10,20,50 and report the averaged metrics for all users in the testing set.
\subsubsection{Recommendation performance comparison using different models.}\hfill 

Table~\ref{tab: performance} reports the performance comparison results, from which we could observe the following and know how to choose the suitable models:
\begin{itemize}
    \item The content-aware methods overall achieve better performance than the CF-based methods(\eg BPR, LightGCN), which indicates the multimodal features always provide richer information about the items and help the recommendation model to reach higher accuracy. Although the recommendation accuracy of VBPR which build based on the BPR has improved a lot. The MMGCN~\cite{wei2019mmgcn} always shows worse performance than LightGCN which does not utilize the multimodal information. It is because of the unreasonable design as LightGCN~\cite{he2020lightgcn} indicated that much noise will influence the performance. MMGCN is the earlier model designed for the multimodal recommendation, At that time the LightGCN technique is not published, but for DualGNN~\cite{wang2021dualgnn}, it adopted the advantage of LightGCN to learn the modality user-item bipartite graph and achieves better performance.
    
    Additionally, the existing content-aware methods largely depend on the representatives of the multimodal features which will cause fluctuating performances among different datasets.\\
    \item Specifically, FREEDOM~\cite{zhou2022tale} achieves the best performance on the baby and sports datasets but gets lower ranks for the food and electronic datasets, which are huge graphs with around 1.7 million interactions. FREEDOM works well on a small graph but it is not suitable for a large graph. ADDVAE always ranks highly except for the FoodRec dataset, which might be because the FoodRec dataset is extracted from the website and the text features for this dataset are messing and the information could hard to be utilized. For the food dataset, there's only a small difference between those models. The performance of SLMRec always depends on the dataset and this model has a long time parameter searching process. LATTICE and BM3 perform stably for the experimental datasets and the performance ranks higher than the CF-based methods and the earlier developed multimodal models like MMGCN and VBPR.\\
    \item The size of the graph and the resource consumed will also influence the choice of algorithms. LATTICE builds an item graph and the model needs many resources to run, which could not be applied to a large graph although it performs well for most datasets. FREEDOM~\cite{zhou2022tale} could achieve high accuracy for the small dataset but the performance applied to large graphs is not the best. It needs fewer computing resources than LATTICE but achieves higher performance. SLMRec could apply to different sizes of graphs but has a long time parameter searching process. We could simplify the search process by reducing the parameters that have been found insensitive by analyzing the results of many datasets. BM3 works well for all sizes of datasets and could achieve the upper rank of performance with fewer resources consumed. 
\end{itemize} 

\begin{table}[htp]
\caption{Performance by different models in terms of Recall and NDCG, Best results are in \textbf{boldface} and the second best is \underline{underlined}.}
\small
\centering
\renewcommand\arraystretch{1.1}{
\begin{tabular}{cccccccc} 
\hline
Dataset & Model & Recall@10 & Recall@20 & Recall@50 & NDCG@10 & NDCG@20 & NDCG@50\\
\hline
\multirow{11}*{Baby} & BPR & 0.0357 & 0.0575 & 0.1054 & 0.0192 & 0.0249 & 0.0345\\
\cline{2-8}
& LightGCN & 0.0479 & 0.0754 & 0.1333 & 0.0257 & 0.0328 & 0.0445 \\
\cline{2-8}
& VBPR & 0.0423 & 0.0663 & 0.1212 & 0.0223 & 0.0284 & 0.0396 \\
\cline{2-8}
& MMGCN & 0.0378 & 0.0615 & 0.1100 & 0.0200 & 0.0261 & 0.0359 \\
\cline{2-8}
& DualGNN & 0.0448 & 0.0716 & 0.1288 & 0.0240 & 0.0309 & 0.0424 \\
\cline{2-8}
& GRCN & 0.0539 & 0.0833 & 0.1464 & 0.0288 & 0.0363 & 0.0490 \\
\cline{2-8}
& LATTICE & 0.0547 & 0.0850 & 0.1477 & 0.0292 & 0.0370 & 0.0497\\
\cline{2-8}
& BM3 & 0.0564 & 0.0883 & 0.1477 & 0.0301 & 0.0383 & 0.0502 \\
\cline{2-8}
& SLMRec & 0.0529 & 0.0775 & 0.1252 & 0.0290 & 0.0353 & 0.0450\\
\cline{2-8}
& ADDVAE & \underline{0.0598} & \underline{0.091} & \underline{0.1508} & \underline{0.0323} & \underline{0.0404} & \underline{0.0525} \\
\cline{2-8}
& FREEDOM & \textbf{0.0627} & \textbf{0.0992} & \textbf{0.1655} & \textbf{0.0330} & \textbf{0.0424} & \textbf{0.0558} \\
\hline
\multirow{11}*{Sports} & BPR & 0.0432 & 0.0653 & 0.1083 & 0.0241 & 0.0298 & 0.0385 \\
\cline{2-8}
& LightGCN & 0.0569 & 0.0864 & 0.1414 & 0.0311 & 0.0387 & 0.0498 \\
\cline{2-8}
& VBPR & 0.0558 & 0.0856 & 0.1391 & 0.0307 & 0.0384 & 0.0492 \\
\cline{2-8}
& MMGCN & 0.0370 & 0.0605 & 0.1078 & 0.0193 & 0.0254 & 0.0350 \\
\cline{2-8}
& DualGNN & 0.0568 & 0.0859 & 0.1392 & 0.0310 & 0.0385 & 0.0493 \\
\cline{2-8}
& GRCN & 0.0598 & 0.0915 & 0.1509 & 0.0332 & 0.0414 & 0.0535 \\
\cline{2-8}
& LATTICE & 0.0620 & 0.0953 & 0.1561 & 0.0335 & 0.0421 & 0.0544\\
\cline{2-8}
& BM3 & 0.0656 & 0.0980 & 0.1581 & 0.0355 & 0.0438 & 0.0561\\
\cline{2-8}
& SLMRec & 0.0663 & 0.0990 & 0.1543 & 0.0365 & 0.0450 & 0.0562\\
\cline{2-8}
& ADDVAE & \underline{0.0709} & \underline{0.1035} & \underline{0.1663} & \textbf{0.0389} & \underline{0.0473} & \underline{0.0600} \\
\cline{2-8}
& FREEDOM & \textbf{0.0717} & \textbf{0.1089} & \textbf{0.1768} & \underline{0.0385} & \textbf{0.0481} & \textbf{0.0618} \\
\hline
\multirow{11}*{FoodRec} & BPR & 0.0303 & 0.0511 & 0.0948 & 0.0188 & 0.0250 & 0.0356 \\
\cline{2-8}
& LightGCN & 0.0331 & 0.0546 & 0.1003 & 0.0210 & 0.0274 & 0.0386 \\
\cline{2-8}
& VBPR & 0.0306 & 0.0516 & 0.0972 & 0.0191 & 0.0254 & 0.0365 \\
\cline{2-8}
& MMGCN & 0.0307 & 0.0510 & 0.0943 & 0.0192 & 0.0253 & 0.0359 \\
\cline{2-8}
& DualGNN & \underline{0.0338} & 0.0559 & \underline{0.1027} & \underline{0.0214} & \underline{0.0280} & \underline{0.0394} \\
\cline{2-8}
& GRCN & \textbf{0.0356} & \textbf{0.0578} & \textbf{0.1063} & \textbf{0.0226} & \textbf{0.0295} & \textbf{0.0411} \\
\cline{2-8}
& LATTICE & 0.0336 & \underline{0.0560} & 0.1012 & 0.0211 & 0.0277 & 0.0388\\
\cline{2-8}
& BM3 & 0.0334 & 0.0553 & 0.0994 & 0.0208 & 0.0274 & 0.0381\\
\cline{2-8}
& SLMRec & 0.0323 & 0.0515 & 0.0907 & 0.0208 & 0.0266 & 0.0362\\
\cline{2-8}
& ADDVAE & 0.0309 & 0.0508 & 0.093 & 0.0186 & 0.0247 & 0.035 \\
\cline{2-8}
& FREEDOM & 0.0333 & 0.0556 & 0.1009 & 0.0212 & 0.0279 & 0.0389 \\
\hline
\multirow{11}*{Elec} & BPR & 0.0235 & 0.0367 & 0.0621 & 0.0127 & 0.0161 & 0.0212 \\
\cline{2-8}
& LightGCN & 0.0363  & 0.0540 & 0.0879 & 0.0204 & 0.0250  & 0.0318  \\
\cline{2-8}
& VBPR & 0.0293 & 0.0458 & 0.0778 & 0.0159 & 0.0202 & 0.0267 \\
\cline{2-8}
& MMGCN & 0.0213 & 0.0343 & 0.0610  & 0.0112 & 0.0146 & 0.0200 \\
\cline{2-8}
& DualGNN & 0.0365 & 0.0542 & 0.0875 & 0.0206 & 0.0252 & 0.0319 \\
\cline{2-8}
& GRCN & 0.0389 & 0.0590 & 0.0970 & 0.0216 & 0.0268 & 0.0345 \\
\cline{2-8}
& LATTICE & - & - & - & - & - & -\\
\cline{2-8}
& BM3 & 0.0437 & 0.0648 & 0.1021 & 0.0247 & 0.0302 & 0.0378\\
\cline{2-8}
& SLMRec & \underline{0.0443} & \underline{0.0651} & \underline{0.1038} & \underline{0.0249} & \underline{0.0303} & \underline{0.0382}\\
\cline{2-8}
& ADDVAE & \textbf{0.0451} & \textbf{0.0665} & \textbf{0.1066} & \textbf{0.0253} & \textbf{0.0308} & \textbf{0.0390} \\
\cline{2-8}
& FREEDOM & 0.0396 & 0.0601 & 0.0998 & 0.0220 & 0.0273 & 0.0353 \\
\hline
\end{tabular}}
\label{tab: performance}
\end{table}

\subsubsection{Recommendation performance comparison using different data split methods.}\hfill

The offline evaluation is based on the historical item ratings or the implicit item feedback. As this method relies on the user-item interactions and the models are all learning based on the supervised signals, we need to split the interactions into train, validation and test sets. There are three main split strategies that we applied to compare the performance: 

\begin{itemize}
\item Random split:\\
As the name suggested, this split strategy randomly selects the train and test boundary for each user. Previously the Leave One Last Item strategy only select the last interaction as the test, but random splitting selects to split the interactions according to the ratio. The disadvantage of the random splitting strategy is that they are not capable to reproduce unless the authors publish how the data split and this is not a realistic scenario without considering the time.
\item User time split:\\
The temporal split strategy splits the historical interactions based on the interaction timestamp by the ratio (e.g. train:validation:test=8:1:1). It split the last percentage of interactions the user made as the test set. VAECF~\cite{liang2018variational}, SVAE~\cite{sachdeva2019sequential} and NGCF~\cite{wang2019neural} are evaluated under this strategy. Although it considers the timestamp, it is still not a realistic scenario because it is still split the train/test sets among all the interactions one user made but did not consider the global time.
\item Global time split:\\
The global time splitting strategy fixed the time point shared by all users according to the splitting ratio. The interactions after the last time point are split as the test set. Additionally, the users of the interactions after the global temporal boundary must be in the training set, which follows the most realistic and strict settings. The limitation of this strategy is that the number of users will be reduced due to deleting the users not existing in the training set.
\end{itemize}
\begin{table}
\caption{Performance by different data split in terms of Recall and NDCG}
\small
\centering
\begin{tabular}{ cccccccc} 
\hline
\multirow{2}*{Dataset} & \multirow{2}*{Model} & random&user time&global time & random&user time&global time\\
\cline{3-8}
~&~&\multicolumn{3}{c}{Recall@10}&\multicolumn{3}{c}{Recall@20}\\
\hline
\multirow{10}*{sports} & MMGCN & 0.0384 & 0.0266 & 0.0140 & 0.0611 & 0.0446 & 0.0245 \\
\cline{2-8}
& BPR & 0.0444 & 0.0322 & 0.0152 & 0.0663 & 0.0509 & 0.0258 \\
\cline{2-8}
& VBPR & 0.0563 & 0.0385 & 0.0176 & 0.0851 & 0.0620 & 0.0298 \\
\cline{2-8}
& DualGNN & 0.0576 & 0.0403 & 0.0181 & 0.0859 & 0.0611 & 0.0297 \\
\cline{2-8}
& GRCN & 0.0604 & 0.0418 & 0.0167 & 0.0915 & 0.0666 & 0.0286 \\
\cline{2-8}
& LightGCN & 0.0568 & 0.0405 & 0.0205 & 0.0863 & 0.0663 & 0.0336 \\
\cline{2-8}
& LATTICE & 0.0641 & 0.0450& 0.0207 & 0.0964 & 0.0699 & 0.0337\\
\cline{2-8}
& BM3 & 0.0646 & 0.0447 & 0.0213 & 0.0955 & 0.0724 & 0.0336 \\
\cline{2-8}
& SLMRec & 0.0651 & 0.0470 & 0.0220 & 0.0985 & 0.0733 & 0.0350\\
\cline{2-8}
& FREEDOM & 0.0708 & 0.0490 & 0.0226 & 0.1080 & 0.0782 & 0.0372 \\
\hline
\multirow{2}*{Dataset} & \multirow{2}*{Model} & \multicolumn{3}{c}{\multirow{2}*{NDCG@10}}& \multicolumn{3}{c}{\multirow{2}*{NDCG@20}}\\
~&~&\multicolumn{3}{c}{~}&\multicolumn{3}{c}{~}\\
\hline
\multirow{10}*{sports}& MMGCN & 0.0202 & 0.0134 & 0.0091 & 0.0261 & 0.0180 & 0.0125 \\
\cline{2-8}
& BPR & 0.0245 & 0.0169 & 0.0102 & 0.0302 & 0.0218 & 0.0135 \\
\cline{2-8}
& VBPR & 0.0304 & 0.0204 & 0.0115 & 0.0378 & 0.0265 & 0.0153 \\
\cline{2-8}
& DualGNN & 0.0321 & 0.0214 & 0.0118 & 0.0394 & 0.0268 & 0.0155 \\
\cline{2-8}
& GRCN & 0.0332 & 0.0219 & 0.0101 & 0.0412 & 0.0282 & 0.0138 \\
\cline{2-8}
& LightGCN & 0.0315 & 0.0220 & 0.0139 & 0.0391 & 0.0286 & 0.0180 \\
\cline{2-8}
& LATTICE & 0.0351 & 0.0238 & 0.0138 & 0.0434 & 0.0302 & 0.0177\\
\cline{2-8}
& BM3 & 0.0356 & 0.0237 & 0.0144 & 0.0436 & 0.0308 & 0.0182 \\
\cline{2-8}
& SLMRec & 0.0364 & 0.0253 & 0.0148 & 0.0450 & 0.0321 & 0.0189\\
\cline{2-8}
& FREEDOM & 0.0388 & 0.0255 & 0.0151 & 0.0485 & 0.0330 & 0.0197 \\
\hline
\end{tabular}
\label{tab: data split}
\end{table}

In order to do a recommendation performance comparison using different data split methods, we did some simple experiments of different data split strategies on the Sports dataset. Table~\ref{tab: data split} shows the performance comparison results in the terms of Recall@k and NDCG@k where k=10,20. And Table~\ref{tab: data split rank} shows the performance ranking of models based on Recall@20 and NDCG@20.

We would like to figure out what impact the data split strategy will make on the state-of-the-art models. Hence we compare the performances of 10 SOTA models on the Sports dataset by applying three data split strategies: Random split, user time split and global time split. As Table~\ref{tab: data split} demonstrates that the models will get different performance numbers by using different data splitting strategies even with the same dataset and evaluation metrics. It is problematic to compare the performance of different models design with different data split strategies and the models are not comparable.

Table~\ref{tab: data split rank} reports the ranks of the 10 SOTA models under different splitting strategies. The rows are sorted by the performance of models under the random splitting strategy, with the up and down arrows indicating the relative rank position swaps compared with random splitting. As we can see, the ranking swaps are observed between the models under different splitting strategies. For example, the GRCN~\cite{wei2020graph} will rank four positions lower under the global time splitting than the random splitting with the performance of BPR, VBPR and LightGCN exceeding. For our experiments, it shows the performance rank of the most effective models and the weakest models will not be influenced, only the rank of middle-performance models has changed under different splitting strategies. Additionally, the time-based splitting will lead to cold start issues of the recommendation systems considered to be solved by leveraging multimodal information. As the performance ranks show, the overall performance of content-based models outperforms the CF-based models over all splitting strategies. However, we get one opposite discovery that there is a rough pattern for the swaps occurred: the CF-based models will be more probably to enhance the ranks under temporal evaluation but some of the content-based models (\eg DualGNN and GRCN) will get lower rank. We guess the reason might be that some of the content-based models did not correctly utilize the multimodal features, which does not help to solve the cold start issue but becomes noise information when facing the cold start issue. 

Most of the multimodal recommender systems use the random splitting strategy, less models use user time splitting and global time splitting. However, the temporal settings are closer to the realistic scenario because the recommender system could not get future data as training data. The robust recommender system should perform stable over all splitting strategies. The behaviors of models depend on how the representations are learned and how the instance has been selected. Hence, we could conclude that the data splitting strategy is the key factor that influences the performance of models and the models for different situations need to choose suitable splitting strategies. However, we only get some superficial inspiration from the simple experiments and we only run 10 models on the sports dataset. If you are interested in the impact of different splitting, you can apply more models with more datasets to find the underlying issues. 

\begin{table}[h]
\caption{Performance Rank of Sports Dataset}
\renewcommand\arraystretch{1.2}{
\begin{minipage}{0.45\linewidth}
\centering
\scalebox{0.65}{
\begin{tabular}{cccc}
\hline
\multirow{2}*{Model}  & \multicolumn{3}{c}{Sports,NDCG@20} \\
\cline{2-4} & random split & User time split & Global time split\\
\hline
MMGCN & 10 &10&10\\
\hline
BPR  & 9 & 9 & 8$\uparrow$1\\
\hline
VBPR&8 &8&7 $\uparrow$1\\
\hline
LightGCN&7 &5$\uparrow$2&4 $\uparrow$3\\
\hline
DualGNN& 6&7$\downarrow$1&6\\
\hline
GRCN &5  &6$\downarrow$1&9 $\downarrow$4\\
\hline
LATTICE& 4 &4 &5$\downarrow$1\\
\hline
BM3&3&3&3\\
\hline
SLMRec&2&2&2\\
\hline
FREEDOM&1&1&1\\
\hline
\end{tabular}}
\end{minipage}
\begin{minipage}{0.45\linewidth}
\centering
\scalebox{0.65}{
\begin{tabular}{cccc}
\hline
\multirow{2}*{Model}  & \multicolumn{3}{c}{Sports,Recall@20} \\
\cline{2-4} & random split & User time split & Global time split\\
\hline
MMGCN & 10&10&10\\
\hline
BPR & 9 & 9 & 9\\
\hline
VBPR&8&7 $\uparrow$1&6  $\uparrow$2 \\
\hline
DualGNN&7&8 $\downarrow$1& 7\\
\hline
LightGCN&6&6&5$\uparrow$1\\
\hline
GRCN&5&5 &8$\downarrow$3\\
\hline
BM3&4&3  $\uparrow$1&4\\
\hline
LATTICE&3&4 $\downarrow$1& 3\\
\hline
SLMRec&2&2&2\\
\hline
FREEDOM&1&1&1\\
\hline
\end{tabular}}
\end{minipage}
\label{tab: data split rank}
}
\end{table}

\subsubsection{Recommendation performance comparison using Different Modalities}\hfill

Multimodal information has been introduced to the recommendation system which helps to alleviate the cold-start problem. Some models utilize the modality information as the side information to enrich the item representations, and some leverage it with the interaction information to enhance the user representation or capture the user preference. 

We are interested in how the modality information benefits the recommendation, and which modality contributes more. We would like to figure out the influence of each modality and provide guidelines for researchers on selecting the modality information utilized.

\begin{figure}
    \centering
    \subfigure[Baby]{
        \includegraphics[width=2.6in]{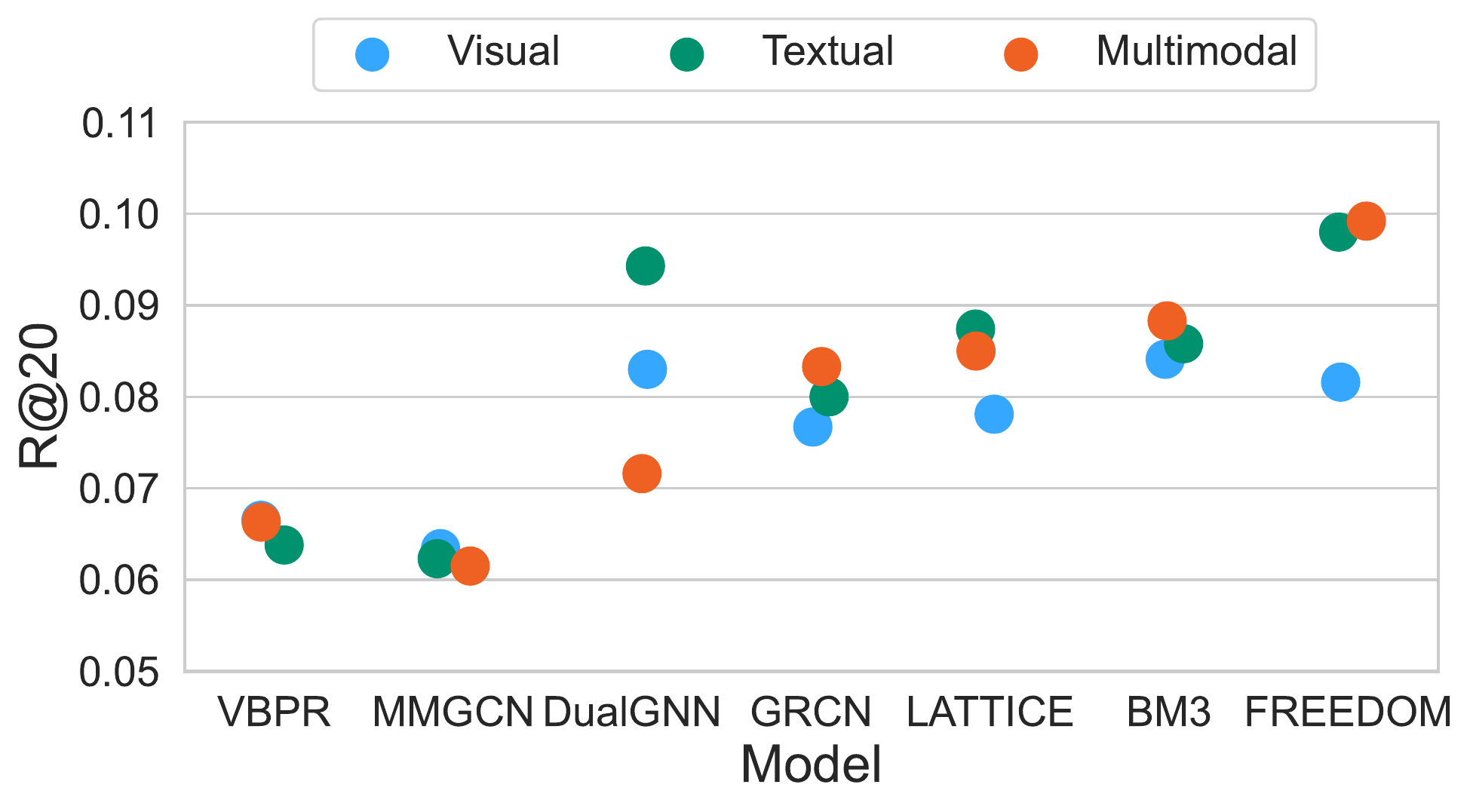}
    }
    \subfigure[Sports]{
	\includegraphics[width=2.6in]{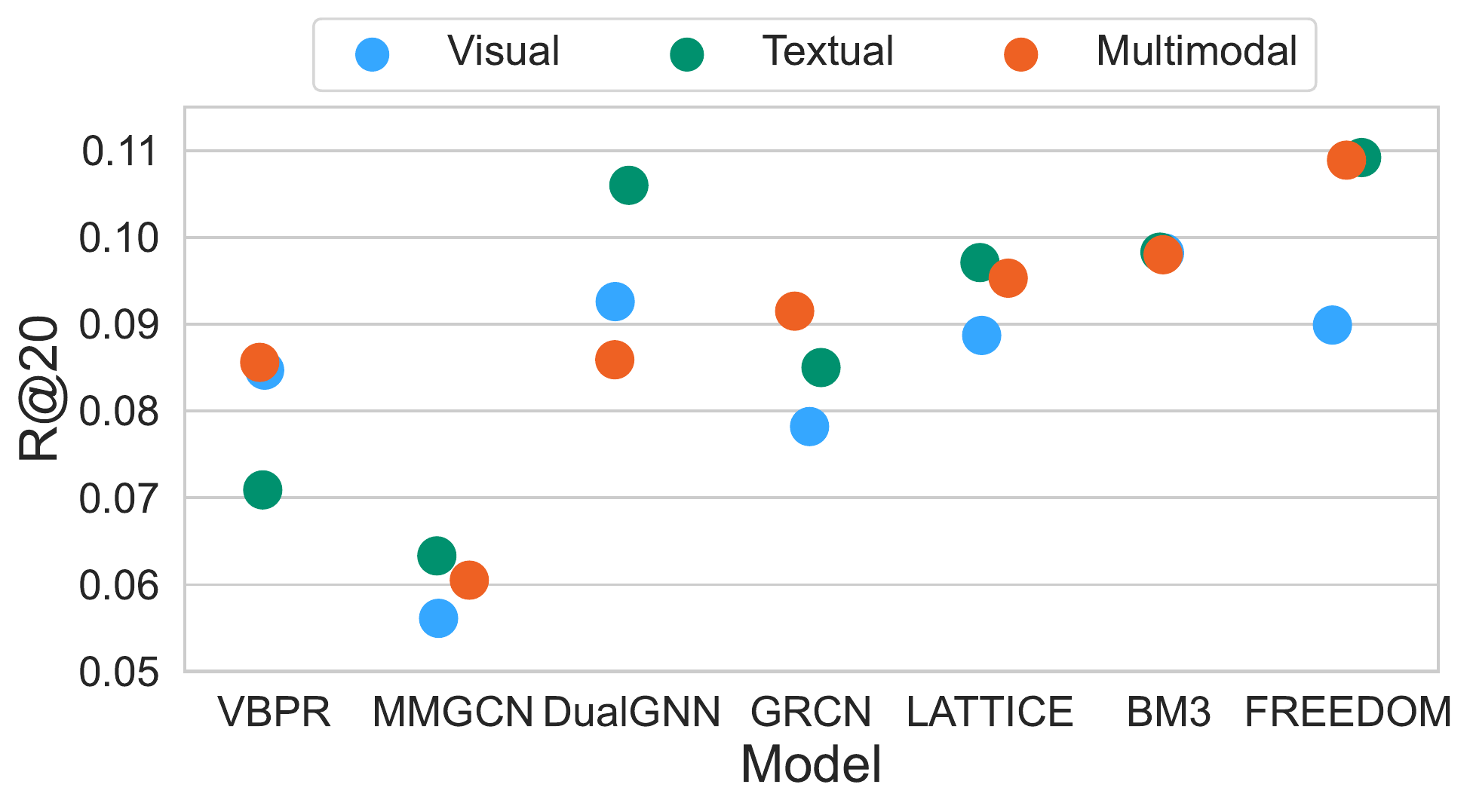}
    }
    \caption{Performance comparison of modalities}
\label{Modality performance}
\end{figure}

We evaluate how the multimodal information influences the recommendation performance of the SOTA models by feeding the single modality information, and compare the performance between using both modalities and the single modality. Table~\ref{tab: text and vision feature recall} shows our experimental results of different modalities. The bold is the best performance of the specific model by feeding other modalities. We draw Fig.~\ref{Modality performance} based on Recall@20 to show the summary and tendency of other modalities which could intuitively show the different influences of modalities on different models. The orange circle represents the performance of multi-modality, the green one represents the performance of textual modality and the blue circle is for visual modality.

We first experiment on the baby and sports datasets. As the result shows, only GRCN~\cite{wei2020graph} with multimodal features outperforms those with a single textual or visual feature over all datasets. For other models, utilizing the single modality information might achieve better performance than fusing the modalities, which indicates that simply fusing the modalities representation by sum, weighted sum and concatenate is not useful to learn the item representation. The simple fusion will destroy the modality-specific information and modality preference learned from each modality. Especially for DualGNN~\cite{wang2021dualgnn}, the textual and visual single modality achieves much better performance than the fusion of modalities, which indicates that the fusion method used influences the quality of representation learned.

Also, we find that the representations are more closely related to the textual modality. The textual modality performance is better or close to the multimodal performance as Fig.~\ref{Modality performance} shows. We are thinking if the dataset influences the result and if those datasets we test are more concentrated on the textual modality. Then we try the Clothing dataset where the visual features are more important, but the result still indicates the importance of textual modality.

For the fusion problem, we need to find out suitable methods which could fuse the multimodal representations with keeping the modality-specific information and capturing the complementary information that a single modality does not contain. The desired result is that the multimodal representations outperform the single modal representations. For the issue that textual modality has better or closely equal performance with the multi-modality, the first reason is the fusion method influences the quality of representations, the second reason might be that the high dimensional visual embedding through an MLP layer to the latent space will compress the data and miss some modality-specific property. Additionally, we found some ablation studies of the baseline models also contain the performance comparison of multi-modality and single-modality~\cite{wei2019mmgcn,wang2021dualgnn} and the multi-modality outperforms single-modality, but the dataset is not published. So the last reason might be because different datasets concentrate on different modalities. The micro video dataset more concentrates on visual modality and it contains more visual modality information to help learn meaningful representations.

\begin{table}
\caption{Performance by different features in terms of Recall}
\small
\centering
\setlength{\tabcolsep}{2.7mm}{
\begin{tabular}{cccccccc} 
\hline
\multirow{2}*{Dataset} & \multirow{2}*{Model} & Both&Textual& Visual&Both&Textual&Visual\\
\cline{3-8}
~&~&\multicolumn{3}{c}{Recall@10}&\multicolumn{3}{c}{Recall@20}\\
\hline
\multirow{8}*{Baby} & VBPR & 0.0423 & 0.0400 & \textbf{0.0428} & 0.0663 & 0.0638 & \textbf{0.0665}\\
\cline{2-8}
& MMGCN & 0.0378 & 0.0365 & \textbf{0.0384} & 0.0615 & 0.0623 & \textbf{0.0634} \\
\cline{2-8}
& DualGNN & 0.0448 & \textbf{0.0612} & 0.0511 & 0.0716 & \textbf{0.0943} & 0.0830 \\
\cline{2-8}
& GRCN & \textbf{0.0539} & 0.0517 & 0.0488 & \textbf{0.0833} & 0.0800 & 0.0767 \\
\cline{2-8}
& LATTICE & \textbf{0.0547} & 0.0546 & 0.0492 & 0.0850 & \textbf{0.0874} & 0.0781\\
\cline{2-8}
& BM3 & 0.0564 & \textbf{0.0571} & 0.0544 & \textbf{0.0883} & 0.0858 & 0.0841 \\
\cline{2-8}
& ADDVAE & - & 0.0598 & - & - & 0.091 & - \\
\cline{2-8}
& FREEDOM & \textbf{0.0627} & 0.0622 & 0.0501 & \textbf{0.0992} & 0.0980 & 0.0816 \\
\hline
\multirow{8}*{Sports} & VBPR & \textbf{0.0558} & 0.0472 & \textbf{0.0558} & \textbf{0.0856} & 0.0709 & 0.0847\\
\cline{2-8}
& MMGCN & 0.0370 & \textbf{0.0401} & 0.0358 & 0.0605 & \textbf{0.0633} & 0.0561 \\
\cline{2-8}
& DualGNN & 0.0568 & \textbf{0.0697} & 0.0615 & 0.0859 & \textbf{0.1060} & 0.0926 \\
\cline{2-8}
& GRCN & \textbf{0.0598} & 0.0545 & 0.0495 & \textbf{0.0915} & 0.0850 & 0.0782 \\
\cline{2-8}
& LATTICE & 0.0620 & \textbf{0.0625} & 0.0572 & 0.0953 & \textbf{0.0971} & 0.0887\\
\cline{2-8}
& BM3 & \textbf{0.0656} & 0.0652 & 0.0649 & 0.0980 & \textbf{0.0983} & 0.0982 \\
\cline{2-8}
& ADDVAE & - & 0.0709 & - & - & 0.1035 & - \\
\cline{2-8}
& FREEDOM & 0.0717	& \textbf{0.0725} & 0.0589 & 0.1089 & \textbf{0.1092} & 0.0899\\
\hline
\multirow{7}*{Clothing} & VBPR & 0.0281 & \textbf{0.0328} & 0.0283 & 0.0415 & \textbf{0.0506} & 0.0416\\
\cline{2-8}
& MMGCN & \textbf{0.0218} & 0.0211 & 0.0203 & \textbf{0.0345} & 0.0342 & 0.0328 \\
\cline{2-8}
& DualGNN & 0.0454 & \textbf{0.0524} & 0.0420 & 0.0683 & \textbf{0.0798} & 0.0636 \\
\cline{2-8}
& GRCN & \textbf{0.0424} & 0.0392 & 0.0351 & \textbf{0.0662} & 0.0612 & 0.0542 \\
\cline{2-8}
& LATTICE & 0.0492 & \textbf{0.0521} & 0.0408 & 0.0733 & \textbf{0.0749} & 0.0614\\
\cline{2-8}
& BM3 & 0.0422 & 0.0413 & \textbf{0.0423} & 0.0621 & 0.0613 & \textbf{0.0626} \\
\cline{2-8}
& FREEDOM & \textbf{0.0629} & 0.0625 & 0.0420 & \textbf{0.0941} & 0.0921 & 0.0640 \\
\hline
\multirow{2}*{Dataset} & \multirow{2}*{Model} & \multicolumn{3}{c}{\multirow{2}*{NDCG@10}}& \multicolumn{3}{c}{\multirow{2}*{NDCG@20}}\\
~&~&\multicolumn{3}{c}{~}&\multicolumn{3}{c}{~}\\
\hline
\multirow{8}*{Baby} & VBPR & 0.0223 & 0.0213 & \textbf{0.0230} & 0.0284 & 0.0275 & \textbf{0.0290}\\
\cline{2-8}
& MMGCN & \textbf{0.0200} & 0.0189 & 0.0197 & \textbf{0.0261} & 0.0256 & 0.0261 \\
\cline{2-8}
& DualGNN & 0.0240 & \textbf{0.0331} & 0.0278 & 0.0309 & \textbf{0.0417} & 0.0360 \\
\cline{2-8}
& GRCN & \textbf{0.0288} & 0.0269 & 0.0258 & \textbf{0.0363} & 0.0342 & 0.0330 \\
\cline{2-8}
& LATTICE & \textbf{0.0292} & 0.0287 & 0.0265 & 0.0370 & \textbf{0.0371} & 0.0339\\
\cline{2-8}
& BM3 & 0.0301 & \textbf{0.0303} & 0.0290 & \textbf{0.0383} & 0.0377 & 0.0367 \\
\cline{2-8}
& ADDVAE & - & 0.0323 & - & - & 0.0404 & - \\
\cline{2-8}
& FREEDOM & \textbf{0.0330} & 0.0325 & 0.0274 & \textbf{0.0424} & 0.0417 & 0.0355 \\
\hline
\multirow{8}*{Sports} & VBPR & \textbf{0.0307} & 0.0256 & 0.0306 & \textbf{0.0384} & 0.0317 & 0.0380\\
\cline{2-8}
& MMGCN & 0.0193 & \textbf{0.0212} & 0.0193 & 0.0254 & \textbf{0.0272} & 0.0245 \\
\cline{2-8}
& DualGNN & 0.0310 & \textbf{0.0379} & 0.0335 & 0.0385 & \textbf{0.0473} & 0.0415 \\
\cline{2-8}
& GRCN & \textbf{0.0332} & 0.0289 & 0.0258 & \textbf{0.0414} & 0.0368 & 0.0332 \\
\cline{2-8}
& LATTICE & 0.0335 & \textbf{0.0336} & 0.0312 & 0.0421 & \textbf{0.0425} & 0.0393\\
\cline{2-8}
& BM3 & \textbf{0.0355} & 0.0352 & 0.0352 & \textbf{0.0438} & 0.0437 & \textbf{0.0438} \\
\cline{2-8}
& ADDVAE & - & 0.0389 & - & - & 0.0473 & - \\
\cline{2-8}
& FREEDOM & 0.0385 & \textbf{0.0387} & 0.0318 & 0.0481 & \textbf{0.0482} & 0.0397 \\
\hline
\multirow{7}*{Clothing} & VBPR & 0.0158 & \textbf{0.0173} & 0.0159 & 0.0192 & \textbf{0.0218} & 0.0193\\
\cline{2-8}
& MMGCN & \textbf{0.0110} & 0.0109 & 0.0107 & \textbf{0.0142} & 0.0142 & 0.0139 \\
\cline{2-8}
& DualGNN & 0.0241 & \textbf{0.0281} & 0.0229 & 0.0299 & \textbf{0.0351} & 0.0283 \\
\cline{2-8}
& GRCN & \textbf{0.0223} & 0.0204 & 0.0178 & \textbf{0.0283} & 0.0260 & 0.0226 \\
\cline{2-8}
& LATTICE & 0.0268 & \textbf{0.0290} & 0.0221 & 0.0330 & \textbf{0.0348} & 0.0273\\
\cline{2-8}
& BM3 & 0.0231 & 0.0225 & \textbf{0.0235} & 0.0281 & 0.0276 &\textbf{ 0.0286} \\
\cline{2-8}
& FREEDOM & \textbf{0.0341} & 0.0336 & 0.0227 & \textbf{0.0420} & 0.0411 & 0.0283 \\
\hline
\end{tabular}
}
\label{tab: text and vision feature recall}
\end{table}

\section{Challenges and Future Research Directions}
Leveraging the multimodal information in the recommender system is highly related to realistic application. The utilization of multimodal content is widespread in recent years to recommend products or related items on e-commerce and social platforms.

In this survey, we have reviewed and classified the most recent and popular used models in the multimodal recommendation area. Particularly, we have introduced a taxonomy of the multimodal recommendation systems by classifying the models according to the techniques they utilized to leverage the multi-modality information that benefits the Recommendation. Additionally, we have introduced the process of multimodal recommendation and listed the common techniques used to do feature extraction, modality fusion and evaluation. Furthermore, we have built a common framework that could help the new researchers easily run the baseline models and know how the recommendation model is designed. The readers could also utilize this framework to easier develop their own models and compare them with baselines. 

During the review and our research process, we find some challenges and possible future research directions.

\subsection{How to effectively fuse multimodal information?}
As the modality information proved to be helpful for the recommendation, how to design the models to better utilized the multi-modality becomes the research challenge. Correctly utilizing multimodal features could improve the performance but use wrongly will make it noisy information. As our experiments show above, the performance of the recommendation system could be improved by utilizing the modality information, but for some models, the single modality will also achieve good performance even higher than utilizing the multimodal information together. We guess for those models, the fusion of multi-modalities causes this issue. Each modality may capture different aspects of the items, we should find a way to fuse them together with keeping the modality-specific information, to learn a multimodal representation containing complementary information that a single modal representation could not include. If the models could fuse the multi-modality features efficiently, the recommendation accuracy should be higher than utilizing the single modality.

Additionally, the modality missing issue is common in the real world. However, some models assume that all modality information is available during training and inference, which will not work when facing the incomplete and missing modality. LRMM~\cite{wang2018lrmm} mitigates the modality missing and cold-start issues by utilizing the generative model to reconstruct the modality-specific embedding and impute the missing modality. 

Future research directions should be studied on how to efficiently utilize the multimodal features: (1) Find out the efficient modality fusion method that can capture complementary information that a single modality can not contain. (2) How to solve the modality missing issue and reconstruct the meaningful representation.

\subsection{How to standardize the data splitting strategy and the general dataset used?}
Several data splitting strategies are applied during the train/test sets splitting process. As the previous experiments show, the models will get different performance numbers by using different data splitting strategies even with the same dataset and evaluation metrics, the ranks of model performances will also be influenced by different splitting strategies. Although the most commonly used is random splitting, temporal-based splitting is closer to the realistic scenario. The robust multimodal recommender models should perform well either with random splitting or time splitting.

The datasets used for multimodal recommendations are not standardized and most of the datasets used in the reviewed papers are not publically available like Kwai, Tiktok and DianPing. Although works on the same public datasets, the setting of datasets is not standardized. Someone would like to do the 5-core setting but others might do the 10-core, which makes the preprocessing of the datasets different. The feature extract techniques will also influence the final representation learned but every paper use different techniques as we showed before.

In future work, we should find out the best preprocessing method with suitable extract techniques and suitable splitting strategies to standardize the experiments' train/test sets.

\subsection{Evaluation metrics}
The evaluation of the recommendation model is an important research topic. The recommender system should not only consider the accuracy but also the other recommendation qualities such as the recommendation list's diversity and the presence of unique items, which may have a significant influence on a recommender system's overall quality~\cite{kaminskas2016diversity}. Additionally, \cite{ge2018evaluation} indicates that the evaluation from the general recommendation might not be totally adopted for the multimodal recommendation. The effectiveness should not just consider accuracy but also the user experience and fairness.

\subsection{Research and application gap}
The multimodal recommendation is not only researched by academic researchers but also by the industrial community. The data used, the information available and even the application scenario are different between them. 
Additionally, the increasing data volume is a big challenge in true-world situations. The scalability and time complexity are critical considerations for choosing models in the industrial community. Industrial research is in a more realistic setting as we said before. Some settings for academic research should be standardized and more practical to develop models suitable for industrial scenarios. The desired recommendation systems should be robust and easily applied to realistic scenarios. More future works should be studied on how to efficiently recommend: (1) Balance the model complexity and scalability when facing large datasets (2) the computation efficiency for high-dimensional tensor and multimodal information.

\subsection{Multimodal sequential recommendation}
A sequential recommendation system is different from the recommendation systems that use collaborative filtering and content-based filtering, as it attempts to understand and model the sequential behaviors of the users over time. The multimodal information could greatly influence the user's preference, however, most of the existing sequential recommendation models ignore such useful information. MML~\cite{pan2022multimodal} incorporates the multimodal side information of items to improve and stabilize the meta-learning process and help to solve the cold-start issue. Therefore, leveraging multimodal information in sequential recommendation systems would be an important direction for future work.

\subsection{Cross domain recommendation}
The cross-domain recommendation system leverage behavior information from other domains or platforms to improve the performance of the target domain. The commonly used technique relies on explicit overlapping data (\eg common users and items) to transfer data across domains. Recently, some works learn the universal representations for users and items that could be applied to the cross-domain recommendation. UniSRec~\cite{hou2022towards} leverages the text information to learn the universal item representations without requiring the common users and items, which could apply to different domains. In the future, leveraging the multimodal information might be able to help models learn the universal representations. 

\section{Conclusion}
In this survey, we provide an extensive review on multimodal recommendation systems. We proposed a clear pipeline for MMRec and list out the commonly used techniques in each step. We classify the models by the learning approaches used for organizing and clustering existing publications. We also make experiments on four commonly used datasets to evaluate the performance of models and provide a common framework for users to easier run the models and develop their models. Additionally, we detail some pressing challenges and promising future research directions. There are a large number of new developed techniques each year, we hope our survey provided a helpful and detailed overview for the researchers, the framework provided an easy and efficient way to run models, and encourage future progress.

\bibliographystyle{ACM-Reference-Format}
\bibliography{ref}

\appendix

\end{document}